%% file: A_Benchmark_for_the_Application_of_Distributed_Control_Techniques_to_the_Electricity_Network_of_the_European_Economic_Area.tex
\documentclass[graybox]{svmult}

\usepackage{type1cm}        
\usepackage{makeidx}         
\usepackage{graphicx}        
\usepackage{multicol}        
\usepackage[bottom]{footmisc}

\usepackage{url}

\usepackage{newtxtext}       %
\usepackage[varvw]{newtxmath}       

\makeindex             

\usepackage{soul}
\usepackage{caption, subcaption} 

\begin{document}

\title*{A Benchmark for the Application of Distributed Control Techniques to the Electricity Network of the European Economic Area }
\titlerunning{The European Economic Area Electricity Network Benchmark}
\author{A. Riccardi, L. Laurenti, and B. De Schutter}
\institute{A. Riccardi \at Delft Center for Systems and Control, Delft University of Technology, Mekelweg 2, 2628 CD Delft, The Netherlands \email{a.riccardi@tudelft.nl}
\and L. Laurenti \at Delft Center for Systems and Control, Delft University of Technology, Mekelweg 2, 2628 CD Delft, The Netherlands \email{l.laurenti@tudelft.nl}
\and B. De Schutter \at Delft Center for Systems and Control, Delft University of Technology, Mekelweg 2, 2628 CD Delft, The Netherlands \email{b.deschutter@tudelft.nl}}
%
%
\maketitle

\abstract*{Each chapter should be preceded by an abstract (no more than 200 words) that summarizes the content. The abstract will appear \textit{online} at \url{www.SpringerLink.com} and be available with unrestricted access. This allows unregistered users to read the abstract as a teaser for the complete chapter.
Please use the 'starred' version of the \texttt{abstract} command for typesetting the text of the online abstracts (cf. source file of this chapter template \texttt{abstract}) and include them with the source files of your manuscript. Use the plain \texttt{abstract} command if the abstract is also to appear in the printed version of the book.}

\abstract{\\
\noindent 
The European Economic Area Electricity Network Benchmark (EEA-ENB) is a multi-area power system representing the European network of transmission systems for electricity to facilitate the application of distributed control techniques. In the EEA-ENB we consider the Load Frequency Control (LFC) problem in the presence of Renewable Energy Sources (RESs), and Energy Storage Systems (ESSs). RESs are known to cause instability in power networks due to their inertia-less and intermittent characteristics, while ESSs are introduced as a resource to mitigate the problem. In the EEA-ENB, particular attention is dedicated to Distributed Model Predictive Control (DMPC), whose application is often limited to small and homogeneous test cases due to the lack of standardized large-scale scenarios for testing, and due to the large computation time required to obtain a centralized MPC action for performance comparison with DMPC strategies under consideration. The second problem is exacerbated when the scale of the system grows. To address these challenges and to provide a real-world-based and control-independent benchmark, the EEA-ENB has been developed. The benchmark includes a centralized MPC strategy providing performance and computation time metrics to compare distributed control within a repeatable and realistic simulation environment.} 
\newpage

\section{Introduction}

\subsection{{The Origin of the Benchmark: Motivation and Challenges}}
The European Economic Area Electricity Network Benchmark (EEA-ENB) is a benchmark designed for the implementation and testing of distributed control strategies for large-scale power networks. The idea behind the benchmark is to build an abstract model of the European network of transmission systems for electricity.  We represent each country of the European economic area as an independent electrical area connected to others through tie lines according to a predefined electricity network topology.  The result is a real-world oriented benchmark that accounts for the presence of renewable generation and Energy Storage Systems (ESSs) in the Load-Frequency Control (LFC) problem of the power network. 
	
The development of the EEA-ENB is essential because no established control model for the European electricity transmission system consistently serves as a reference for distributed control techniques, especially with energy storage systems and renewable energy sources. Additionally, the use case for the EEA-ENB is not restricted only to the pure development of control strategies. With minimal modifications, it can also be used for other applications, such as the economic optimization of network operation, the study of network expansion strategies, testing of security and privacy features, and simulation of emergency situations such as cascading blackouts and network restoration.

To assess the time and computation requirements for the implementation of a distributed control strategy we implement centralized Model Predictive Control (MPC) on the network. Together with the value of the cost function of centralized MPC developed, this provides the user metrics to evaluate the advantages and disadvantages in the implementation of a specific distributed control technique. The EEA-ENB is formulated with a modular approach such that extensions can be implemented if needed, allowing for various application scenarios as mentioned before. The stability of the network is assessed through the study of LFC problem. Moreover, another application that is particularly relevant for this benchmark is the economical optimization of energy trading among network agents. The EEA-ENB can also be employed to formulate Distributed MPC (DMPC) techniques in the presence of hybrid dynamics thanks to a modified ESS dynamics reported. Additional extensions, not included in this work, include the characterization of each electrical area according to the deregulated energy market through the modeling of generation plants, the auction system for scheduling energy production across the various generation companies, and the market of power exchanges between different electrical areas \cite{bevrani_RobustPowerSystem_2014}.

The main challenge in controlling the EEA-ENB has to be sought in its scale: 26 electrical areas are considered, each subject to distinct variations in load requests and renewable generation. When using a growing number of control agents the computation time of a centralized control action becomes increasingly prohibitive, thus, distributed control approaches are required.
	
\subsection{Load Frequency Control in Modern Power Networks}
The LFC problem is a crucial challenge in power systems, and it has a particular socio-economic interest  \cite{kundur_PowerSystemStability_2022}. The LFC problem gained interest in the research community in the 1970s \cite{fosha_MegawattFrequencyControlProblem_1970} after some major systems events led to cascading blackouts \cite{haesalhelou_SurveyPowerSystem_2019}. These problems typically arise when unexpected changes in the load of a power system occur, with consequent shifts in the operating frequency of the electrical area under consideration, and the propagation of this effect to neighboring areas. In the last decades cascading blackouts have been exacerbated by the increasing diffusion of renewable energy sources, which are posing new challenges for LFC of interconnected power grids due to their intermittent and stochastic nature, and inertia-less generation \cite{ranjan_LiteratureSurveyLoad_2022}.

Nowadays, new strategies to increase network robustness are constantly sought \cite{bevrani_RobustPowerSystem_2014}. This is the reason why ESSs are fundamental in modern energy grids: they allow for more efficient use of energy, optimizing its usage based on the demand, and they can be used to counteract the inertia-less properties of renewable energy sources. Therefore, part of the modeling section of this chapter is dedicated to ESSs, from the simplest dynamical formulation to more complex hybrid formulations.

Formally, the main control problem solved in the EEA-ENB is the regulation to zero of the frequency deviation of the network from the nominal value. This problem is solved in the presence of unexpected changes in the load, renewable generation, and ESSs. Early approaches to the solution were mainly based on PID control theory. With the progression of technology, more advanced techniques have been implemented, such as variable gain scheduling, fuzzy logic control, artificial neural networks, and optimal control \cite{kundur_PowerSystemStability_2022, ranjan_LiteratureSurveyLoad_2022}.

In this chapter, we propose MPC as reference control technique for the benchmark, and DMPC as its natural extension. The choice of MPC is related to the fact that it provides the optimal control action according to a certain cost function defined by the user, while incorporating constraints on the evolution of the state and control. Nevertheless, the EEA-ENB is designed to be control-independent, and virtually all control techniques can be implemented on it. For a detailed list of control approaches for the LFC problem, we refer to \cite{ranjan_LiteratureSurveyLoad_2022}.

\section{Problem Description} 

\subsection{System Description} \label{subsec:System_Description}
\begin{figure}[tbp!]
	\centering
	\includegraphics[width=.7\linewidth]{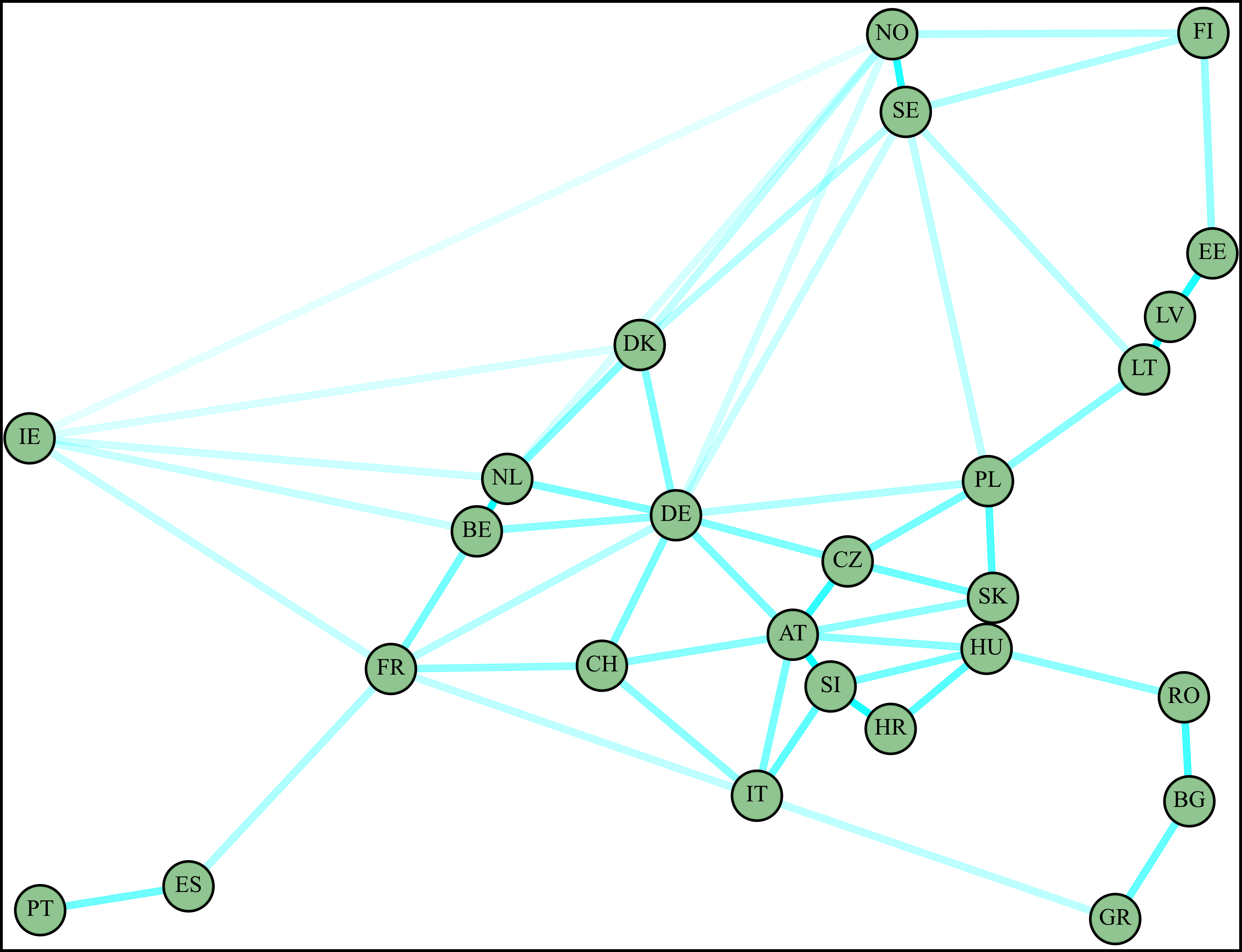}
	\caption{\color{black}Electrical topology of the EEA-ENB. Each node represents an electrical area, coinciding with a single country, whereas each edge is a transmission line. The transparency of the edges represents the strength of interaction among the areas.}
	\label{fig:Topology26_Topology}
\end{figure}

The EEA-ENB is composed of 26 interacting electrical areas connected through tie lines and uses real-world data acquired from the European Network of Transmission System Operators for Electricity (ENTSO-E) transparency platform accessible from \cite{_ENTSOE}. Each area represents an equivalent electrical machine aggregating the inertia and dispatchable capacity of generators in that specific area, a modeling technique commonly used in the context of LFC \cite{kundur_PowerSystemStability_2022}. The electrical topology of the network is derived from the grid map also provided by ENTSO-E \cite{_ENTSOE}. The benchmark is constituted by 26 control areas due to considerations about the availability and scale of the data about the 31 members of the EEA. The electrical topology of the resulting network is reported in Fig.\ \ref{fig:Topology26_Topology}, where each country is labeled with the respective ISO code. Positions of the areas in the space are selected according to their geographical centroids, and, on this basis, the lengths of the tie lines are defined using the Euclidean distance as reported in Table \ref{tab:distance}. 
\begin{table}[tbp!]
	\caption{Lengths of the tie lines expressed in $10^3$ [km].}
	\centering
	\resizebox{\textwidth}{!}{ 
\begin{tabular}{c|cccccccccccccccccccccccccc}
	& AT & BE & BG & HR & CZ & DK & EE & FI & FR & DE & GR & HU & IE & IT & LV & LT & NL & NO & PL & PT & RO & SK & SI & ES & SE & CH \\
	\hline        
	AT &0 & 0 & 0 & 0 & 2.65 & 0 & 0 & 0 & 0 & 4.82 & 0 & 5.61 & 0 & 4.76 & 0 & 0 & 0 & 0 & 0 & 0 & 0 & 5.88 & 1.85 & 0 & 0 & 5.58 \\ 
	BE &0 & 0 & 0 & 0 & 0 & 0 & 0 & 0 & 4.69 & 5.77 & 0 & 0 & 13.19 & 0 & 0 & 0 & 1.75 & 0 & 0 & 0 & 0 & 0 & 0 & 0 & 0 & 0 \\ 
	BG&0 & 0 & 0 & 0 & 0 & 0 & 0 & 0 & 0 & 0 & 4.02 & 0 & 0 & 0 & 0 & 0 & 0 & 0 & 0 & 0 & 3.01 & 0 & 0 & 0 & 0 & 0 \\ 
	HR&0 & 0 & 0 & 0 & 0 & 0 & 0 & 0 & 0 & 0 & 0 & 3.61 & 0 & 0 & 0 & 0 & 0 & 0 & 0 & 0 & 0 & 0 & 2.12 & 0 & 0 & 0 \\ 
	CZ&2.65 & 0 & 0 & 0 & 0 & 0 & 0 & 0 & 0 & 5.13 & 0 & 0 & 0 & 0 & 0 & 0 & 0 & 0 & 4.67 & 0 & 0 & 4.33 & 0 & 0 & 0 & 0 \\ 
	DK&0 & 0 & 0 & 0 & 0 & 0 & 0 & 0 & 0 & 5.03 & 0 & 0 & 17.82 & 0 & 0 & 0 & 5.44 & 11.56 & 0 & 0 & 0 & 0 & 0 & 0 & 10.22 & 0 \\ 
	EE&0 & 0 & 0 & 0 & 0 & 0 & 0 & 6.37 & 0 & 0 & 0 & 0 & 0 & 0 & 2.2 & 0 & 0 & 0 & 0 & 0 & 0 & 0 & 0 & 0 & 0 & 0 \\ 
	FI&0 & 0 & 0 & 0 & 0 & 0 & 6.37 & 0 & 0 & 0 & 0 & 0 & 0 & 0 & 0 & 0 & 0 & 8.99 & 0 & 0 & 0 & 0 & 0 & 0 & 8.89 & 0 \\ 
	FR&0 & 4.69 & 0 & 0 & 0 & 0 & 0 & 0 & 0 & 9.35 & 0 & 0 & 12.38 & 11.19 & 0 & 0 & 0 & 0 & 0 & 0 & 0 & 0 & 0 & 8.58 & 0 & 6.09 \\ 
	DE&4.82 & 5.77 & 0 & 0 & 5.13 & 5.03 & 0 & 0 & 9.35 & 0 & 0 & 0 & 0 & 0 & 0 & 0 & 4.98 & 15.23 & 9.06 & 0 & 0 & 0 & 0 & 0 & 13.41 & 4.84 \\ 
	GR&0 & 0 & 4.02 & 0 & 0 & 0 & 0 & 0 & 0 & 0 & 0 & 0 & 0 & 10.94 & 0 & 0 & 0 & 0 & 0 & 0 & 0 & 0 & 0 & 0 & 0 & 0 \\ 
	HU&5.61 & 0 & 0 & 3.61 & 0 & 0 & 0 & 0 & 0 & 0 & 0 & 0 & 0 & 0 & 0 & 0 & 0 & 0 & 0 & 0 & 5.87 & 1.48 & 4.63 & 0 & 0 & 0 \\ 
	IE&0 & 13.19 & 0 & 0 & 0 & 17.82 & 0 & 0 & 12.38 & 0 & 0 & 0 & 0 & 0 & 0 & 0 & 13.84 & 27.51 & 0 & 0 & 0 & 0 & 0 & 0 & 0 & 0 \\ 
	IT&4.76 & 0 & 0 & 0 & 0 & 0 & 0 & 0 & 11.19 & 0 & 10.94 & 0 & 0 & 0 & 0 & 0 & 0 & 0 & 0 & 0 & 0 & 0 & 3.81 & 0 & 0 & 5.84 \\ 
	LV&0 & 0 & 0 & 0 & 0 & 0 & 2.2 & 0 & 0 & 0 & 0 & 0 & 0 & 0 & 0 & 1.69 & 0 & 0 & 0 & 0 & 0 & 0 & 0 & 0 & 0 & 0 \\ 
	LT&0 & 0 & 0 & 0 & 0 & 0 & 0 & 0 & 0 & 0 & 0 & 0 & 0 & 0 & 1.69 & 0 & 0 & 0 & 5.55 & 0 & 0 & 0 & 0 & 0 & 10.14 & 0 \\ 
	NL&0 & 1.75 & 0 & 0 & 0 & 5.44 & 0 & 0 & 0 & 4.98 & 0 & 0 & 13.84 & 0 & 0 & 0 & 0 & 16.99 & 0 & 0 & 0 & 0 & 0 & 0 & 0 & 0 \\ 
	NO&0 & 0 & 0 & 0 & 0 & 11.56 & 0 & 8.99 & 0 & 15.23 & 0 & 0 & 27.51 & 0 & 0 & 0 & 16.99 & 0 & 0 & 0 & 0 & 0 & 0 & 0 & 2.28 & 0 \\ 
	PL&0 & 0 & 0 & 0 & 4.67 & 0 & 0 & 0 & 0 & 9.06 & 0 & 0 & 0 & 0 & 0 & 5.55 & 0 & 0 & 0 & 0 & 0 & 3.37 & 0 & 0 & 10.93 & 0 \\ 
	PT&0 & 0 & 0 & 0 & 0 & 0 & 0 & 0 & 0 & 0 & 0 & 0 & 0 & 0 & 0 & 0 & 0 & 0 & 0 & 0 & 0 & 0 & 0 & 4.34 & 0 & 0 \\ 
	RO&0 & 0 & 3.01 & 0 & 0 & 0 & 0 & 0 & 0 & 0 & 0 & 5.87 & 0 & 0 & 0 & 0 & 0 & 0 & 0 & 0 & 0 & 0 & 0 & 0 & 0 & 0 \\ 
	SK&5.88 & 0 & 0 & 0 & 4.33 & 0 & 0 & 0 & 0 & 0 & 0 & 1.48 & 0 & 0 & 0 & 0 & 0 & 0 & 3.37 & 0 & 0 & 0 & 0 & 0 & 0 & 0 \\ 
	SI&1.85 & 0 & 0 & 2.12 & 0 & 0 & 0 & 0 & 0 & 0 & 0 & 4.63 & 0 & 3.81 & 0 & 0 & 0 & 0 & 0 & 0 & 0 & 0 & 0 & 0 & 0 & 0 \\ 
	ES&0 & 0 & 0 & 0 & 0 & 0 & 0 & 0 & 8.58 & 0 & 0 & 0 & 0 & 0 & 0 & 0 & 0 & 0 & 0 & 4.34 & 0 & 0 & 0 & 0 & 0 & 0 \\ 
	SE&0 & 0 & 0 & 0 & 0 & 10.22 & 0 & 8.89 & 0 & 13.41 & 0 & 0 & 0 & 0 & 0 & 10.14 & 0 & 2.28 & 10.93 & 0 & 0 & 0 & 0 & 0 & 0 & 0 \\ 
	CH&5.58 & 0 & 0 & 0 & 0 & 0 & 0 & 0 & 6.09 & 4.84 & 0 & 0 & 0 & 5.84 & 0 & 0 & 0 & 0 & 0 & 0 & 0 & 0 & 0 & 0 & 0 & 0 \\ 
\end{tabular}}
\label{tab:distance}
\end{table}
In this graph representation, each node is associated with dynamics incorporating generation, storage, consumption, and interaction behaviors of the considered electrical area and of its neighborhood. In particular, an electrical area is composed of a multiplicity of autonomous subsystems working together to guarantee the satisfaction of the setpoints assigned by the area-level controller. The aggregation of those subsystems allows one to define an equivalent electrical machine for each area. Specifically, each area may comprehend the following: 
\begin{itemize}
	\item A \textit{dispatchable generator} used to model all sources of energy that can be actively controlled to balance the load. Conventional power sources are hydroelectric turbines, nuclear power plants and gas, oil, or coil turbines. Those sources are associated with an aggregated power generation that we can allocate at each time step according to the production limits of each area.
	\item \textit{Non-dispatchable generation} associated with renewable energy production, such as wind and solar generation, which have intermittent and stochastic nature. We assume that data are available both for the exact value of the produced power and for day-ahead forecasts.
	\item An \textit{ESS} used to accumulate and supply energy at the best convenience and according to the control strategy implemented. In general, energy storage systems can be classified into three macro-categories: electrical storage (e.g. ultracapacitors), electrochemical storage (e.g. batteries), and mechanical storage (e.g. water reservoirs). However, this distinction is not considered in the benchmark, but it is suggested as a possible extension. Following the same approach used for dispatchable generation, we consider the aggregated storage and power of all the ESSs in the electrical area.
	\item A \textit{load demand} for which measurements and day-ahead forecasts are available. 
\end{itemize} 
Those components contribute to the internal load-frequency balance of the electrical area. Moreover, a power exchange among areas is present over the tie lines reported in the electrical topology. This interaction must also be accounted for in the overall power balance. 

\subsection{System Dynamics}\label{subsec:System_Dynamics}
\begin{figure}[tbp!]
	\centering
	\includegraphics[width=\linewidth]{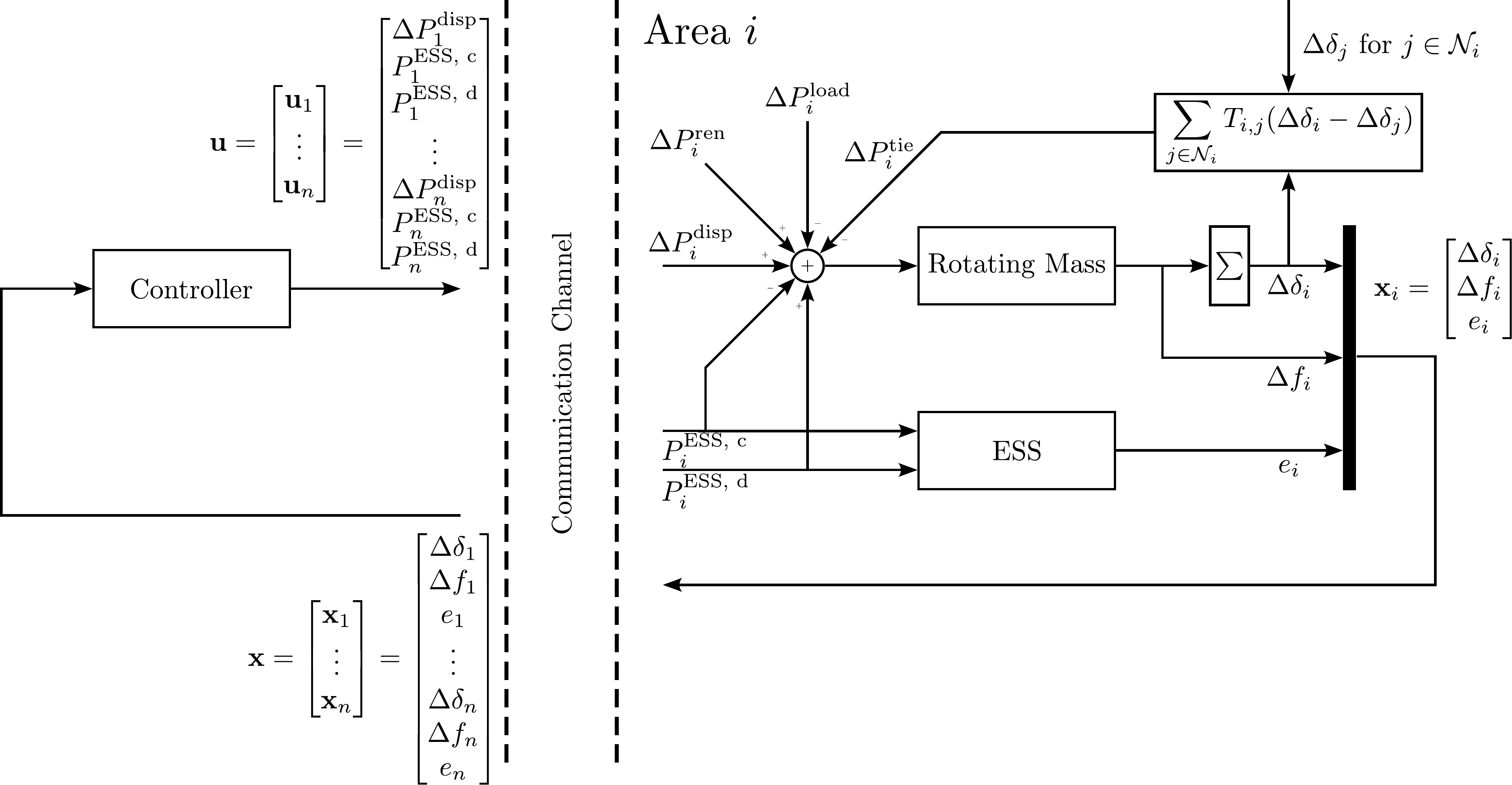}
	\caption{\color{black}Schematic of the network of equivalent machines, with details of the $i$-th electrical area. }
	\label{fig:controlscheme}
\end{figure}

The topology of the power system is represented as a graph $\mathcal{G} = (\mathcal{V},\mathcal{E})$ where each node $v_i\in\mathcal{V}$ is associated with an independent electrical area $i$, and each undirected edge $\epsilon_{ij} = \epsilon_{ji} = (v_i,v_j) \in\mathcal{E} \subseteq \mathcal{V}\times\mathcal{V}$ is a tie line connecting adjacent areas $i$ and $j$, allowing for bidirectional power flow. In our case, we have 26 nodes, one for each electrical area. The presence of an edge represents the existence of a power connection. 
For each node $v_i$, we define its neighborhood as $\mathcal{N}_i = \{v_j \in \mathcal{V}\,|\,(v_i,v_j)\in\mathcal{E}\}$, i.e.\ the set of nodes connected to the node $v_i$.
To each node $v_i$, $i=1,\ldots, 26$ an equivalent electrical machine is associated according to the schematic in Fig.\ \ref{fig:controlscheme}. Each electrical area $i$ is always characterized by at least three states: the angle $\delta_i$ [deg] of the rotor, the operating frequency $f_i$ [Hz] of the equivalent machine, and the energy $e_i$ [GWh] stored in the ESS. The control inputs for the $i$-th area are the deviation $\Delta P_i^{\text{disp}}$ [GW] of dispatchable power production w.r.t.\ the scheduled value, and the power $P_i^{\text{ESS, c}}$ [GW] supplied to or $P_i^{\text{ESS, d}}$ [GW] withdrawn from the ESS. Additionally, each area is subjected to the influence of external inputs: the variation in the load request $\Delta P_i^{\text{load}}$ [GW], renewable energy production $\Delta P_i^{\text{ren}}$ [GW], and the power transmitted over the tie lines $\Delta P_i^{\text{tie}}$ [GW] connected to area $i$. 

For this system, it is common to assume \cite{bevrani_RobustPowerSystem_2014, kundur_PowerSystemStability_2022} a linearized discrete-time model around an operating point $(\delta_{0,i}, f_{0,i})$ for the power angle and frequency dynamics for each area $i$. For all $i\in\{1,\ldots,26\}$, we assume $\delta_{0,i} = 30$ [deg], but this depends on the scheduled power exchanges among electrical areas as specified later, with limits $\delta_0 \in (0,90)$ [deg]; moreover, the operating frequency of the European power network is $f_{0,i} = 50$ [Hz] \cite{kundur_PowerSystemStability_2022}. Regarding the ESS of the $i$-th area, the simplest model capturing the charging and discharging characteristics of a storage system is the linear representation also reported in \cite{sioshansi_EnergyStorageModelingStateoftheArt_2022}. Extensions of this model, and alternative formulations of the ESS dynamics, as the PWA description in \cite{parisio_ModelPredictiveControl_2014b}, are discussed in Section \ref{subsubsec:ESS}. For the tie lines interaction, we also use a linearized equation \cite{kundur_PowerSystemStability_2022} under the assumption that machine angle deviations are small enough, which will be guaranteed through operating constraints in the control formulation. 

To summarize, the state, input, and external input of the $i$-th area are the vectors:
\begin{equation}
	\mathbf{x}_i = \begin{bmatrix}
		\Delta \delta_i & \Delta f_i & e_i
	\end{bmatrix}^\intercal \quad 
	\mathbf{u}_i = \begin{bmatrix}
		\Delta P_i^{\text{disp}} & P_i^{\text{ESS, c}} & P_i^{\text{ESS, d}}
	\end{bmatrix}^\intercal \quad 
	\mathbf{w}_i = \begin{bmatrix}
		\Delta P_i^{\text{load}} & \Delta P_i^{\text{ren}} & \Delta P_i^{\text{tie}}
	\end{bmatrix}^\intercal,
\end{equation}
and their aggregation provides the respective definition of the state, input, and external input vectors for the overall network:
\begin{equation}
	\mathbf{x} = \begin{bmatrix}
		\mathbf{x}_1^\intercal & \cdots & \mathbf{x}_{26}^\intercal
	\end{bmatrix}^\intercal \quad 
	\mathbf{u} = \begin{bmatrix}
		\mathbf{u}_1^\intercal & \cdots & \mathbf{u}_{26}^\intercal
	\end{bmatrix}^\intercal \quad 
	\mathbf{w} = \begin{bmatrix}
		\mathbf{w}_1^\intercal & \cdots & \mathbf{w}_{26}^\intercal
	\end{bmatrix}^\intercal. 
\end{equation}
Assuming that the discrete-time dynamics obtained through forward Euler discretization has sampling time $\tau$, the dynamics of the $i$-th electrical area has the form:
\begin{flalign} \label{eq:overall_dynamics}
	& \mathcal{S}_i:\left\{ \begin{matrix}
		\Delta \delta_i(k+1) = \Delta \delta_i(k) + \tau 2 \pi \Delta f_i(k) \hfill \\
	\begin{matrix}
		\Delta f_i(k+1) = \left(1-\frac{\tau}{T_{\text{p}, i}}\right) \Delta f_i(k) + \tau \frac{K_{\text{p}, i}}{T_{\text{p}, i}}\big(\Delta P_i^{\text{disp}}(k)-\Delta P_i^{\text{load}}(k) + \Delta P_i^{\text{ren}}(k) \\  \hfill   - \Delta P_i^{\text{tie}}(k) - P_i^{\text{ESS, c}}(k) + P_i^{\text{ESS, d}}(k)\big)
	\end{matrix} \\
	e_i(k+1) = e_i(k) + \tau\left(\eta_{i}^{\text{c}} P_{i}^{\text{ESS, c}}(k) - \frac{1}{\eta_{i}^{\text{d}}}P_{i}^{\text{ESS, d}}(k)\right) \hfill 
	\end{matrix}
	\right. \\
	& \Delta P_i^{\text{tie}}(k) = \sum_{j\in\mathcal{N}_i} T_{ij}(\Delta \delta_i(k) - \Delta \delta_j(k)),
\end{flalign}
where $K_{\text{p}, i}$ and $T_{\text{p}, i}$ are respectively the gain and the time constants of the dynamics of the rotating mass; $\eta_{i}^{\text{c}}$ and $\eta_{i}^{\text{d}}$ are charging and discharging rates of the battery with $0<\eta_{i}^{\text{c}},\eta_{i}^{\text{d}}<1$;  and $T_{ij}$ [GW/deg] is the gain associated with the tie line $(i,j)$, i.e. $T_{ij} = k_{ij}/d_{ij}$, which depends on the geographical distance $d_{ij}$ [km] among the electrical areas, and on the gain $k_{ij}$ [km$\cdot$GW/deg], which is assumed to be equal to 1 for all $i$, $j$ in this chapter.

\subsection{Assumptions and Operating Conditions} \label{subsec:Assumptions}
The electrical angle deviation is constrained as $-30 \leq \Delta \delta_i \leq 30$, so that the electrical angle satisfies $0 \leq \delta_i \leq 60$, with $\delta_{0,i} = 30$ [deg]. For the operating frequency we assume the range $-0.04 \leq \Delta f_i \leq 0.04$, with $f_{0,i} = 50$ [Hz] \cite{bevrani_RobustPowerSystem_2014, kundur_PowerSystemStability_2022}. For the ESSs, we consider the maximum storage capacity to be equal to the total dispatchable capacity, i.e.\ $0 \leq e_i \leq e_{i, \text{max}}^{\text{disp}}$, for each area $i$, with $e_{i, \text{max}}^{\text{disp}} = 1\cdot P_{i, \text{max}}^{\text{disp}}$ [GWh].
For each electrical area $i$ the following state constraints hold:
\begin{equation} \label{eq:states_limits}
	\begin{matrix}
		-30 \leq \Delta \delta_i \leq 30 &\qquad&
		-0.04 \leq \Delta f_i \leq 0.04 &\qquad&
		0 \leq e_i \leq P_{i, \text{max}}^{\text{disp}} 
	\end{matrix}
\end{equation}
Input limits are selected such that the total available dispatchable or storage capacity can be allocated over one hour:
\begin{equation} \label{eq:input_limits}
	- \frac{P_{i, \text{max}}^{\text{disp}}}{1440} \leq \Delta P_i^{\text{disp}} \leq \frac{P_{i, \text{max}}^{\text{disp}}}{1440} \qquad 	0 \leq P_i^{\text{ESS, c}}, \,\, P_i^{\text{ESS, d}} \leq \frac{P_{i, \text{max}}^{\text{disp}}}{1440}
\end{equation}
The sampling time of the system is $\tau = 2.5$ [s], which is 10 times faster than the time constant $T_{\text{p}, i} = 25$ [s]. A variation in the external inputs occurs every $1440$ time steps, i.e.\ every hour. A simulation of $24\cdot1440 = 34560$ steps would use $24$ hours of real-world data about load and renewable generation, see Section \ref{subsec:data} for additional details.

\subsection{Extensions and Alternative Formulations}
\label{subsec:extensions}
We propose three directions to modify or extend the proposed dynamics: a PWA formulation of the ESS dynamics, an extension to include the behavior of turbines and pumps, and an augmented state representation to describe the energy market. Other possible extensions are reported at the end of this section.

\subsubsection{ESS Hybrid Dynamics} \label{subsubsec:ESS}
Assuming that the ESSs can only be in a charging or discharging state at each time step, their dynamics can be described with the following Piecewise Affine Linear (PWA) equations \cite{parisio_ModelPredictiveControl_2014b}:
\begin{equation}\label{eq:ESS_hybrid}
	e_i(k+1) = \left\{\begin{matrix}
		e_i(k) + \tau \eta_{i}^{\text{c}} \Delta P_i^{\text{ESS}}(k) & \text{if} & \Delta P_i^{\text{ESS}}(k) \geq 0 \hfill \\ 
		e_i(k) + \tau \frac{1}{\eta_{i}^{\text{d}}} \Delta P_i^{\text{ESS}}(k) & \text{if} & \Delta P_i^{\text{ESS}}(k) < 0
	\end{matrix}
	\right.
\end{equation}
In this formulation, the charging and discharging inputs used in \eqref{eq:overall_dynamics} are substituted by a single input $P_i^{\text{ESS}}$, representing the total power exchange of the electrical area with the ESS. This formulation is completely different from the linear one in \eqref{eq:overall_dynamics}, and it can be demonstrated that the two representations are equivalent only if the ESS is lossless, i.e.\ is if $\eta^{\text{c}}=\eta^{\text{d}}=1$. 

\subsubsection{Turbine and Pump Dynamics Extension}
A finer representation of the system would include the presence of a turbine for the generation of the dispatchable power, and of a turbine/pump system for mechanical ESS to allocate and use energy in the water reservoirs (this is not necessary for other types of ESSs). This concept can be applied both to the ESS formulation in \eqref{eq:overall_dynamics} and \eqref{eq:ESS_hybrid}. Additional states are introduced in this new description: the signals previously considered in \eqref{eq:overall_dynamics} and \eqref{eq:ESS_hybrid} as inputs are now the states of the turbines or pump. Additionally, new inputs $u_i^{\text{disp}} $, $u_i^{\text{ESS, c}}$, $u_i^{\text{ESS, d}}$ are introduced to control the turbines and pump. Specifically, if we consider the linear formulation \eqref{eq:overall_dynamics} for the $i$-th electrical area we have: 
\begin{equation}
	\mathbf{x}_i = \begin{bmatrix}
		\Delta \delta_i & \Delta f_i & e_i & \Delta P_i^{\text{disp}}  &  P_i^{\text{ESS, c}} & P_i^{\text{ESS, d}}
	\end{bmatrix}^\intercal \quad 
	\mathbf{u}_i = \begin{bmatrix}
		u_i^{\text{disp}} & u_i^{\text{ESS, c}} & u_i^{\text{ESS, d}} 
	\end{bmatrix}^\intercal 
\end{equation}
and the dynamics \eqref{eq:overall_dynamics} is augmented with the update equations:
\begin{equation}
\begin{matrix}
		\Delta P_i^{\text{disp}}(k+1) = \left(1 - \frac{\tau}{T_{\text{t},i}}\right)\Delta P_i^{\text{disp}}(k) + \tau \frac{K_{\text{t},i}}{T_{\text{t},i}} u_i^{\text{disp}} \hfill \\
		P_i^{\text{ESS, c}}(k+1) = \left(1 - \frac{\tau}{T_{\text{c},i}}\right)\Delta P_i^{\text{ESS, c}}(k) + \tau \frac{K_{\text{c},i}}{T_{\text{c},i}} u_i^{\text{ESS, c}} \hfill \\
		P_i^{\text{ESS, d}}(k+1) = \left(1 - \frac{\tau}{T_{\text{d},i}}\right)\Delta P_i^{\text{ESS, d}}(k) + \tau \frac{K_{\text{d},i}}{T_{\text{d},i}} u_i^{\text{ESS, d}} \hfill
\end{matrix}
\end{equation}
where $T_{\text{t},i}$, $T_{\text{c},i}$, $T_{\text{d},i}$ and $K_{\text{t},i}$, $K_{\text{c},i}$, $K_{\text{d},i}$ are respectively the time constants and gains of the turbine and storage turbine/pump of the $i$-th electrical area. As good engineering practice, the time constants $T_{\text{t},i}$, $T_{\text{c},i}$, $T_{\text{d},i}$ are selected to be at least 10 times smaller than $T_{\text{p},i}$, and accordingly the sampling time $\tau$ has to be at least 100 times smaller than the original one in \eqref{eq:overall_dynamics}, i.e.\ $\tau = 0.025$ [s]. For further details see \cite{bevrani_RobustPowerSystem_2014, ersdal_ModelPredictiveLoadFrequency_2016}.

\subsubsection{State Augmentation and Total Production Constraints} \label{subsubsec:energy_market}
{ An aspect not usually considered in LFC systems is the total dispatchable production limit. In this benchmark, using real-world data, we want to constrain the total dispatchable production to be non-negative, and smaller than the overall capacity of a certain area.  To this end, the dynamics \eqref{eq:overall_dynamics} are augmented with the two states $P_i^{\text{disp}}$, $P_i^{\text{tie}}$, which evolve according to:
\begin{equation}
	\begin{matrix}
		P_i^{\text{disp}}(k+1) = P_i^{\text{disp}}(k) + \tau \Delta P_i^{\text{disp}}(k) \hfill \\
		P_i^{\text{tie}}(k+1) = P_i^{\text{tie}}(k) + \tau \Delta P_i^{\text{tie}}(k) \hfill
	\end{matrix}
\end{equation}
In this way, we can also impose limits on the overall dispatchable generation for each electrical area according to the data acquired from \cite{_ENTSOE}. Specifically, each area is subjected to $0 \leq P_i^{\text{disp}}(k) \leq P_{i, \text{max}}^{\text{disp}}$. Some areas may have a renewable production that exceeds the total load request. Thus, this constraint ensures that the excess is stored for later use or transmitted to neighboring areas through machine angles adjustments. As initial condition, we assume to have a dispatchable production that compensates for the total load request and that accounts for the renewable generation:
\begin{equation}
	P_{i}^{\text{disp}}(0) = \max\left\{0; P_{i}^{\text{load}}(0) - P_{i}^{\text{ren}}(0)\right\}.
\end{equation}
This choice of $P_{i}^{\text{disp}}(0)$ is to guarantee it to be positive, thus we select it as the maximum between zero and the difference $P_{i}^{\text{load}}(0) - P_{i}^{\text{ren}}(0)$.
}

\subsubsection{Additional Extensions}
The equivalent machine modeling approach can also be applied to the deregulated energy market \cite{bevrani_RobustPowerSystem_2014}. This approach involves defining various actors for electricity production in different regions, each with its dispatchable generation capacity. These actors, known as generation companies, can be represented by individual turbines that aggregate the inertia of all the generators within the same company. Additionally, there are ESSs that aggregate the storage capacities of each company. Thus, in each area $i$, there is a certain number of dispatchable generators and ESSs. 
A centralized auction system determines which generation company supplies energy to each area, considering cross-border production, electrical topology, and predefined operational strategies.
 
We also highlight the fact that each electrical area can be further subdivided into frontier sectors and a central sector, with tie-lines connecting them. This subdivision of the electrical topology can be used for energy trade modeling. The central sector may account for the generation of critical infrastructures and is connected to all the frontier sectors. Each frontier sector is connected to the frontier sector of a neighboring area and to the central sector of the area it belongs to. This further subdivision of the topology can be used to mitigate the effect of power transmission from adjacent areas, to ensure enhanced stability of the central sector, and to define scheduled power transmissions among neighboring areas. 

Future research should also consider the exploration of different ESS technologies, the challenges related to their implementation, their feasibility, and economic sustainability, all aspects that can contribute to the further refinement of the EEA-ENB.

\subsection{Goal of the Control System}
{ The main control goals for the benchmark are the following:
\begin{itemize}
	\item \textit{Regulation} of the frequency deviation $\Delta f_i = f_i - f_{i, 0}$ of each electrical area to zero, so that the frequency of the network stays at the desired value $f_0 = 50$ [Hz]. This is also the main goal of the control system. Moreover, we require to regulate the machine angle deviation $\Delta \delta_i = \delta_i - \delta_{i,0}$ to zero, such that the efficiency of the machine is preserved. 
	\item \textit{Operational constraints satisfaction}. In addition to the regulation goals, the control system should also ensure that the operational constraints of Section \ref{subsec:Assumptions} and \ref{subsubsec:energy_market} are satisfied. Ensuring that these constraints are satisfied is of primary importance for the stability of the network, and its correct functioning. Similar constraints should be enforced also for the augmented models in Section \ref{subsec:extensions}.
	\item \textit{Disturbance rejection}. From a control perspective, the external signals of load and renewable generation variations in \eqref{eq:overall_dynamics} can be interpreted as disturbances to reject. Moreover, in the benchmark both measurements and forecasts for these signals are provided, allowing the user to represent different operational scenarios. 
	\item \textit{Minimization of the control effort}. The control inputs of the benchmark are the variation in dispatchable generation and power exchange with the ESS for each area $i$, namely $\mathbf{u}_i = [\begin{matrix}
		\Delta P_i^{\text{disp}} & P_i^{\text{ESS, c}} & P_i^{\text{ESS, d}}
	\end{matrix}]^\intercal$. Minimizing the values of these quantities while ensuring the correct functioning of the system is another relevant feature for control design. The control effort is quantified as the vector product $\mathbf{u}_i^{\intercal}\mathbf{u}_i$ for each area $i$.
\end{itemize}
Other control objectives can be designed depending on the study that wants to be conducted on the benchmark. For example, if an economic MPC problem is formulated using electricity prices, then the total monetary cost for running the network can be considered, defined as $\text{\texteuro}\cdot\text{MW}$ for each agent, for each energy source, and at each time step. With this approach, the least expensive network operation strategy can be defined, trading off the lower operational cost of the network with its stability. In this regard, if soft constraints are implemented to limit the frequency deviation, then the total-time-spent and the average-time-spent outside the optimal operation interval of the frequency can also be considered as a performance indicator. Moreover, other control goals can be considered that are more specific to the technological implementation of the network. Those could regard the number of charging and discharging cycles of the batteries, their average charge level, or also the electrical machine angle deviation w.r.t.\ the most efficient one. This means that we might incorporate operational and maintenance costs in the benchmark and consider them as a way to compare control strategies.}

\section{Benchmark Design} 
\subsection{Input Data} \label{subsec:data}
\begin{figure}[tbp!] 
	\begin{minipage}{.85\textwidth}
		\begin{subfigure}[t]{0.48\textwidth}
			\includegraphics[width=\textwidth]{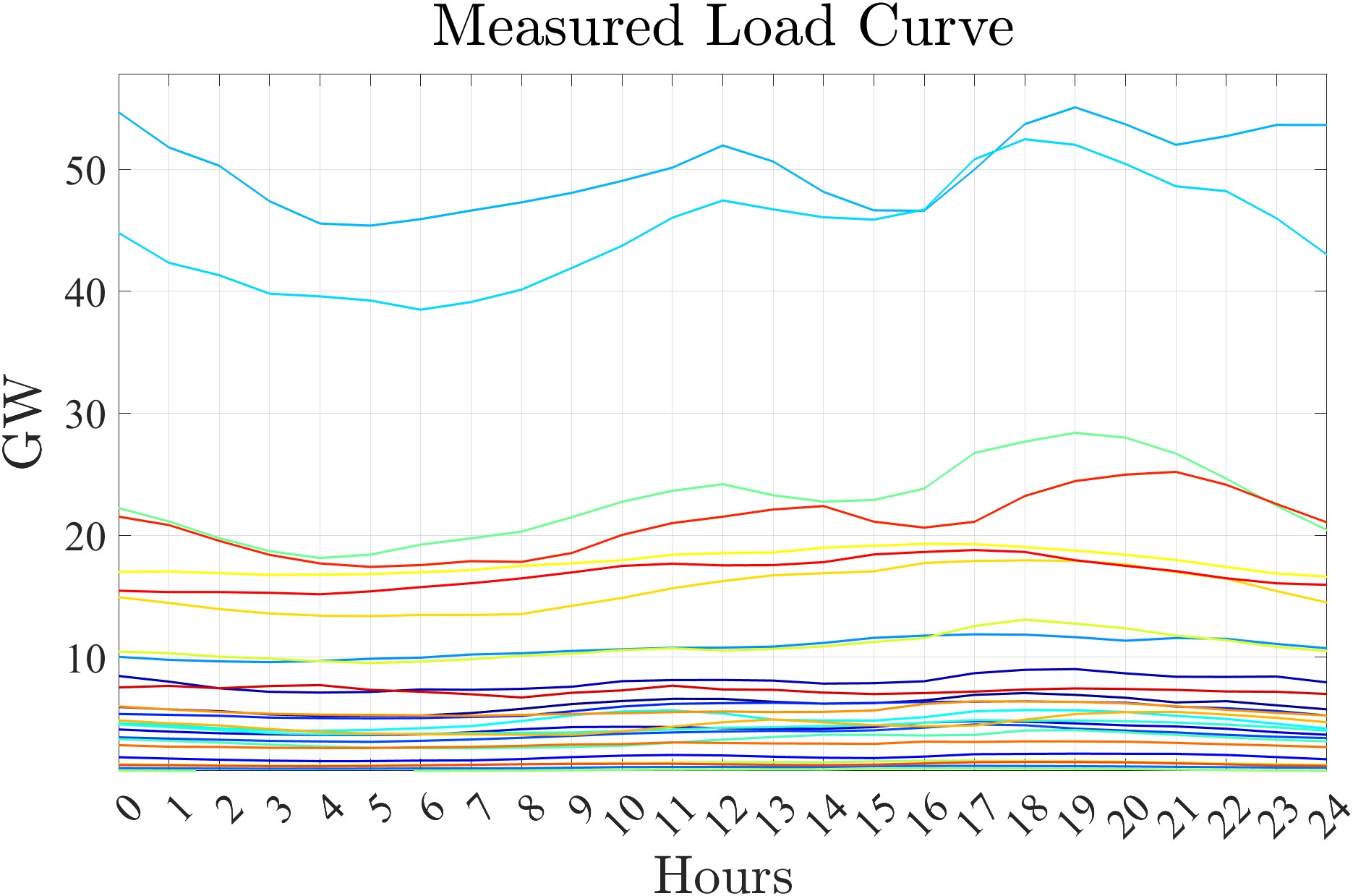}
		\end{subfigure} \hfill
		\begin{subfigure}[t]{0.48\textwidth}
			\includegraphics[width=\textwidth]{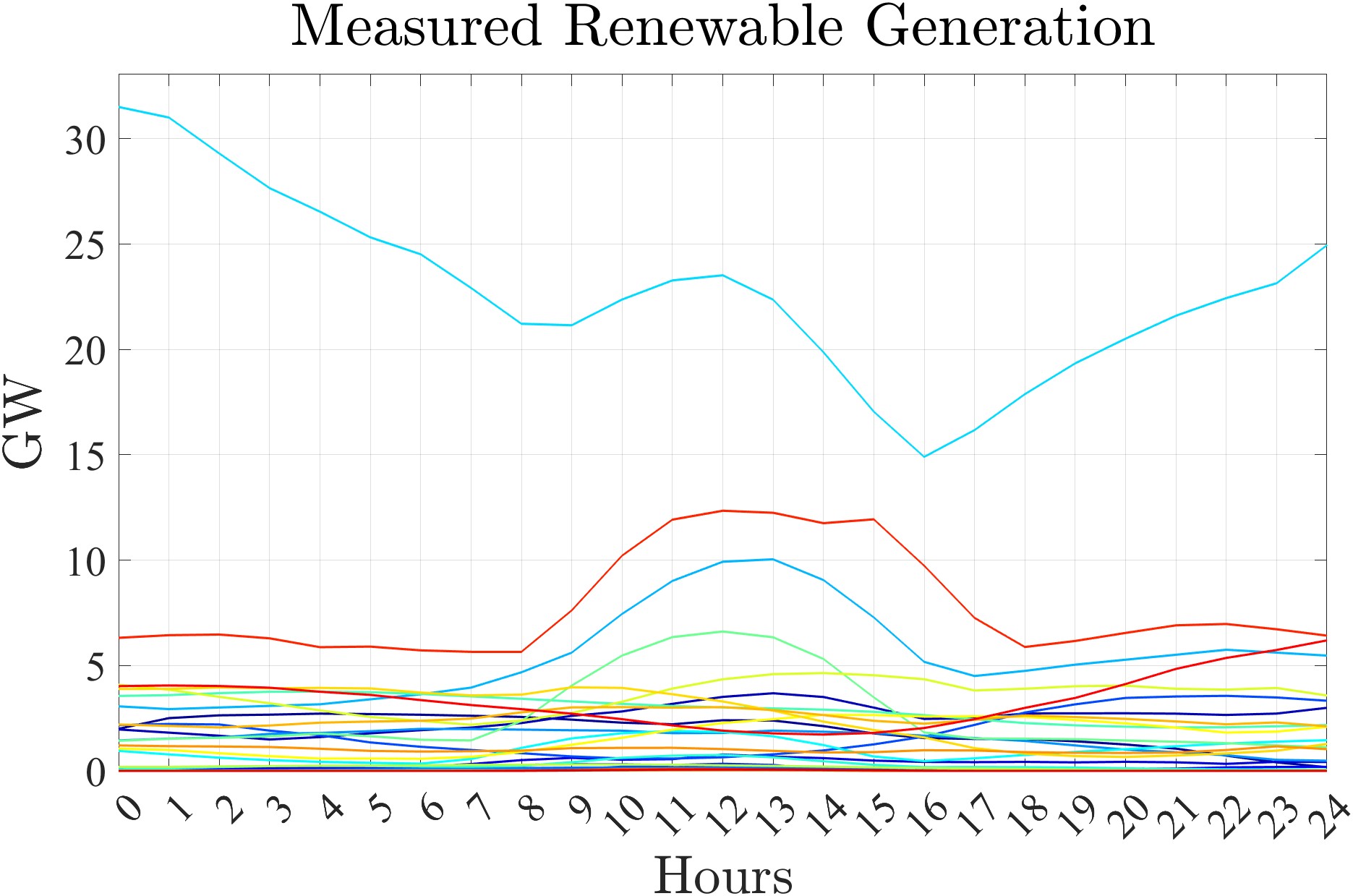}
		\end{subfigure} \hfill \vspace{\baselineskip}\\
		\begin{subfigure}[t]{0.48\textwidth}
			\includegraphics[width=\textwidth]{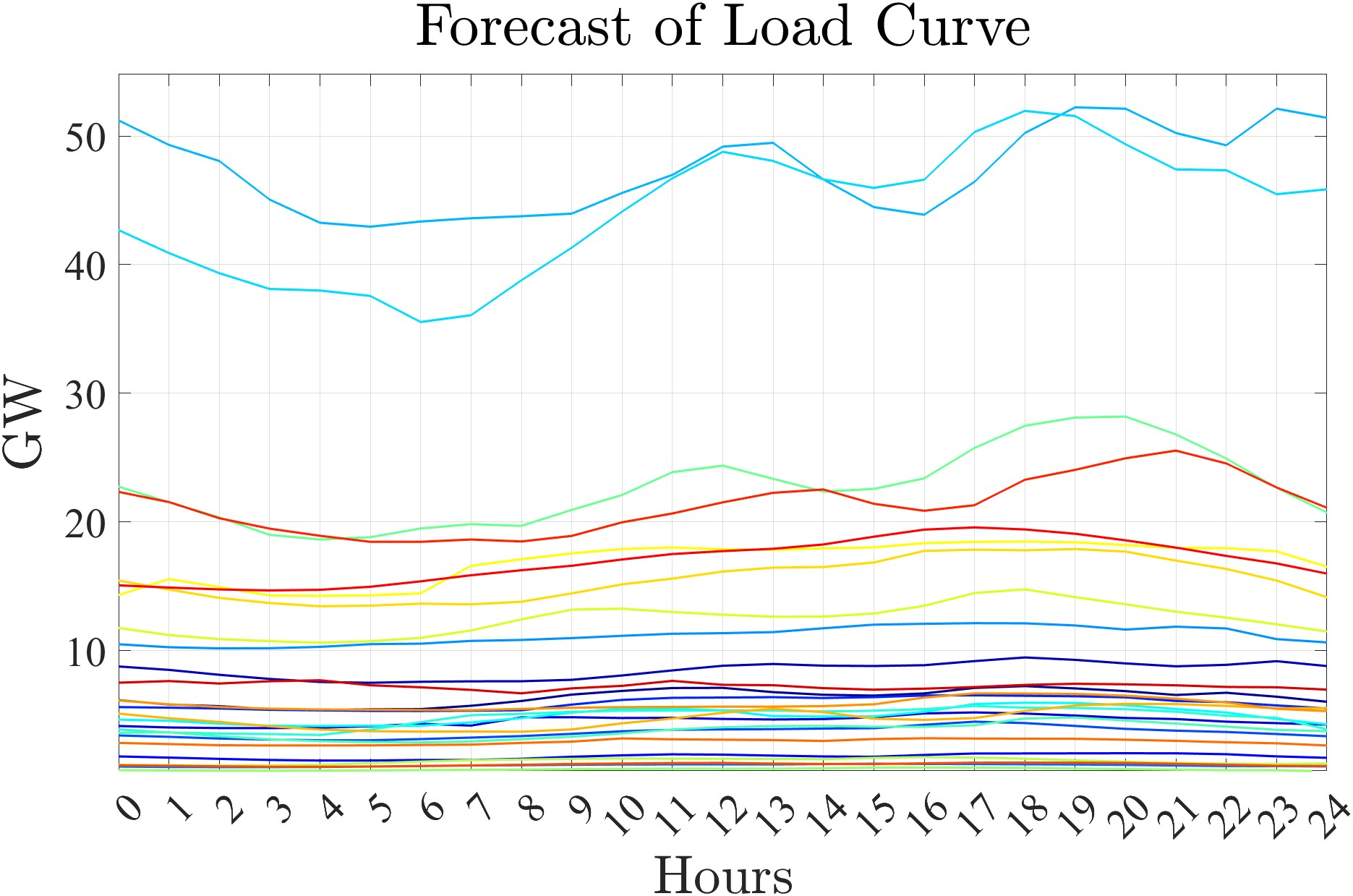}
		\end{subfigure} \hfill
		\begin{subfigure}[t]{0.48\textwidth}
			\includegraphics[width=\textwidth]{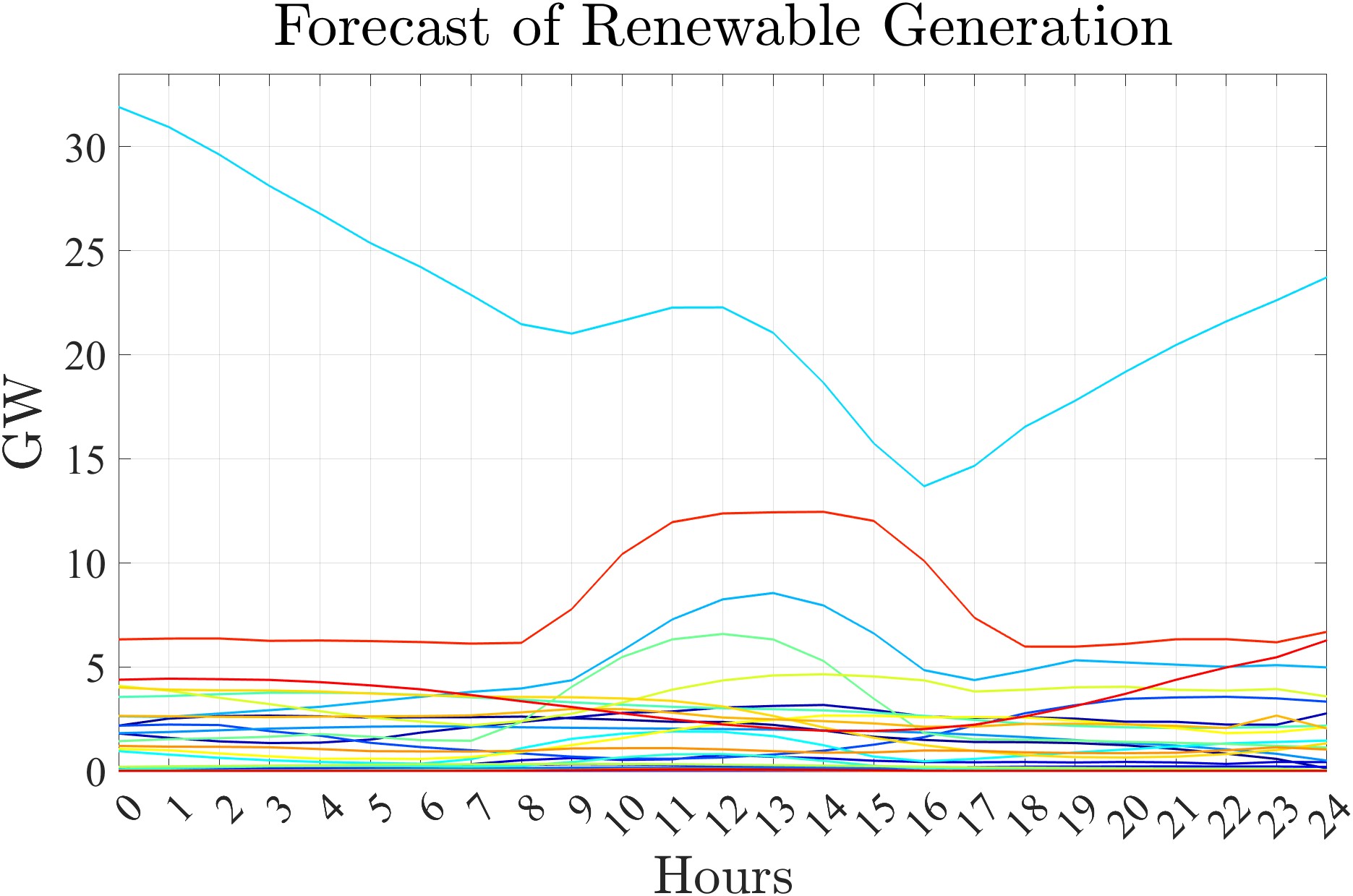}
		\end{subfigure} \hfill
	\end{minipage}%
	\hspace{0.03\linewidth}
	\begin{minipage}{0.11\textwidth}
		\includegraphics[width=\textwidth]{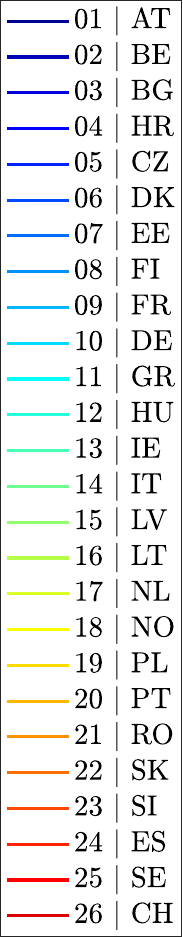}
	\end{minipage}
	\caption{Measurements and forecasts of the load request and renewable generation of each electrical area.}
	\label{fig:data}
\end{figure}

The network of equivalent machines is modeled using real data about load requests, renewable generation, and dispatchable capacities of the 26 European states selected for the implementation. Data is acquired from the ENTSO-E electricity transparency platform \cite{_ENTSOE}. As an example, data for the 24 hours of January 1, 2022, are reported in Fig.\ \ref{fig:data}.
Raw data is available every hour for each area considered, both for measurements and day-ahead forecasts. 

\subsection{Implementation Details}
\begin{figure}[tbp!] 
	\begin{minipage}{.85\textwidth}
		\begin{subfigure}[t]{0.48\textwidth}
			\includegraphics[width=\textwidth]{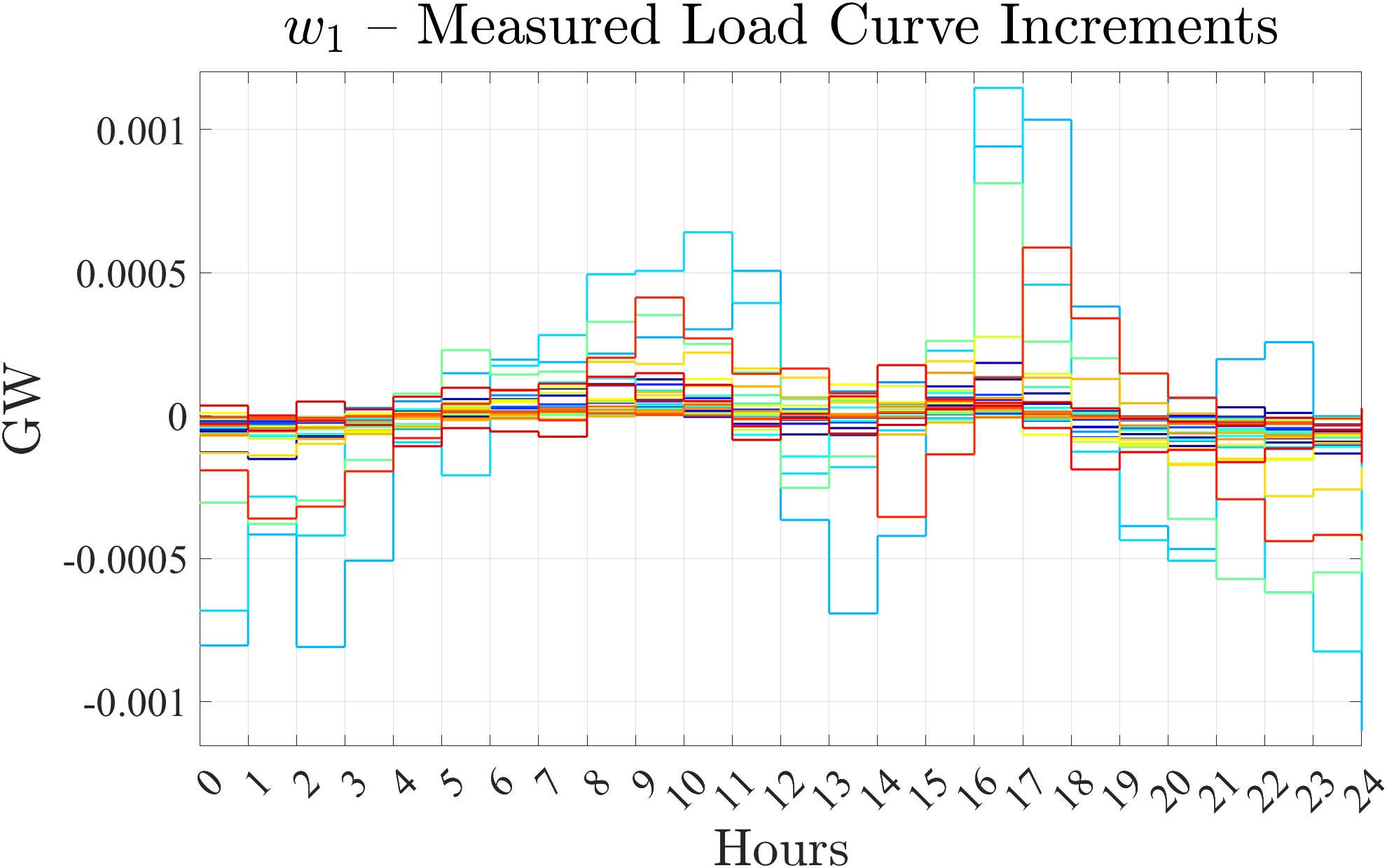}
		\end{subfigure} \hfill
		\begin{subfigure}[t]{0.48\textwidth}
			\includegraphics[width=\textwidth]{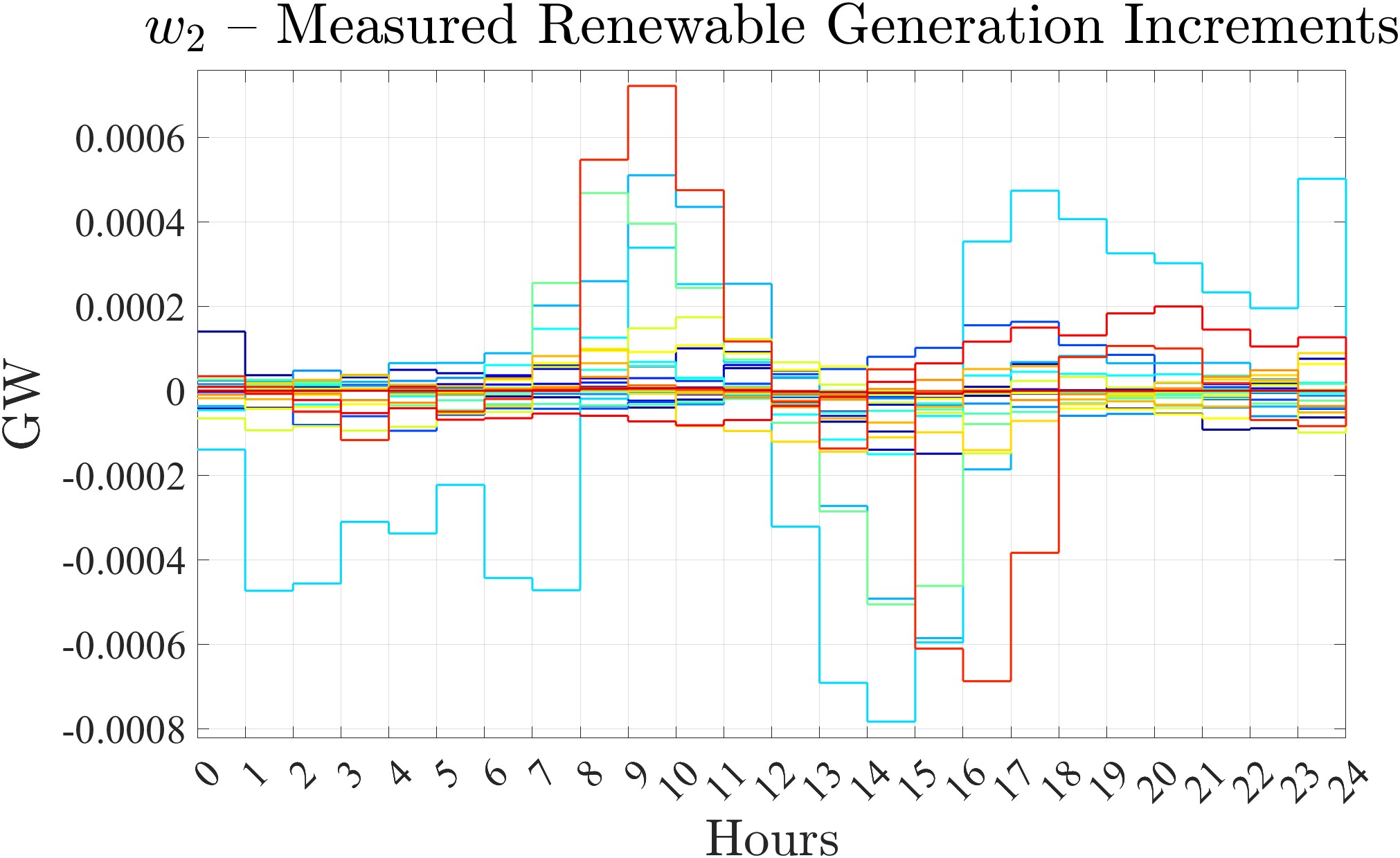}
		\end{subfigure} \hfill \vspace{\baselineskip}\\
		\begin{subfigure}[t]{0.48\textwidth}
			\includegraphics[width=\textwidth]{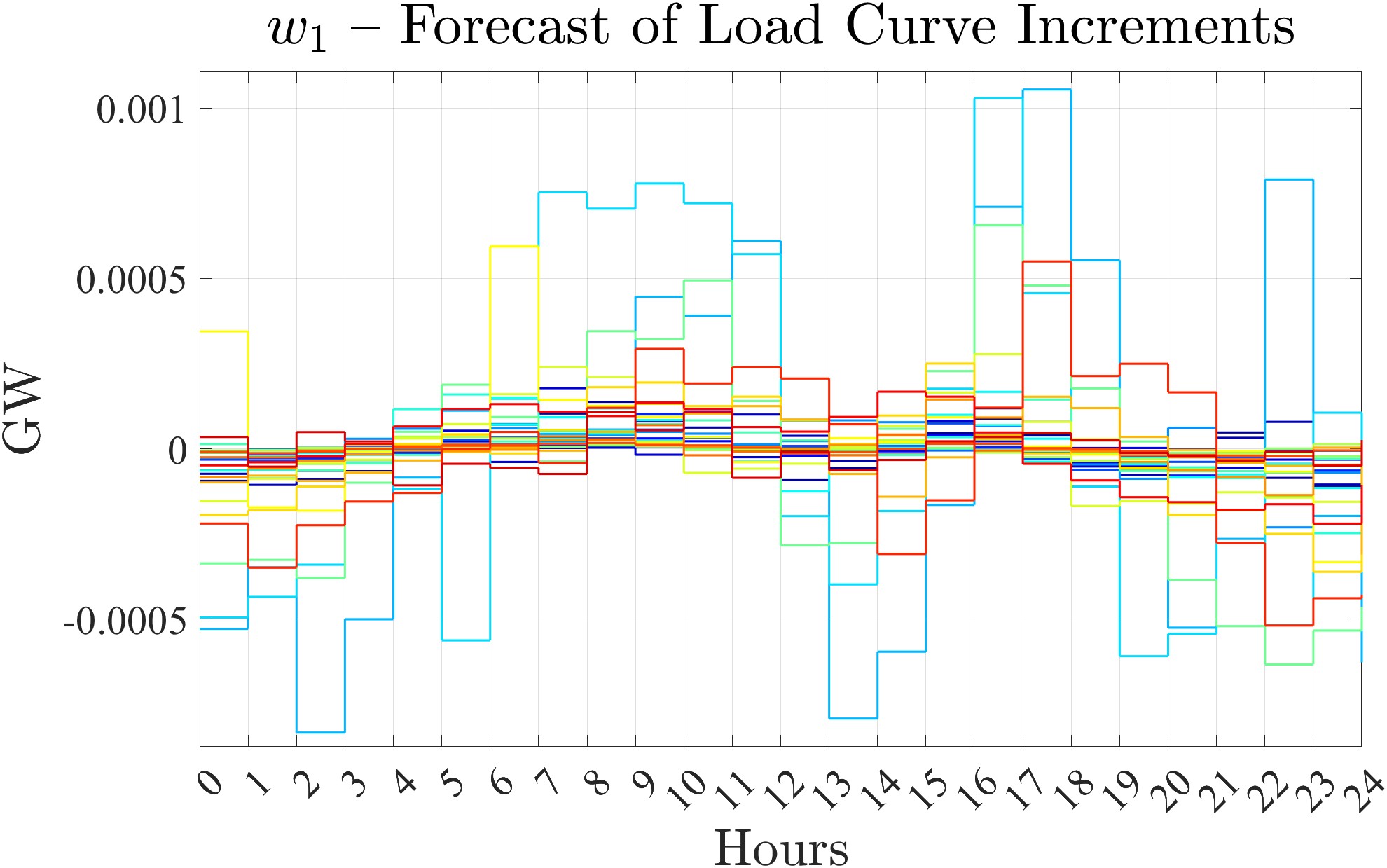}
		\end{subfigure} \hfill
		\begin{subfigure}[t]{0.48\textwidth}
			\includegraphics[width=\textwidth]{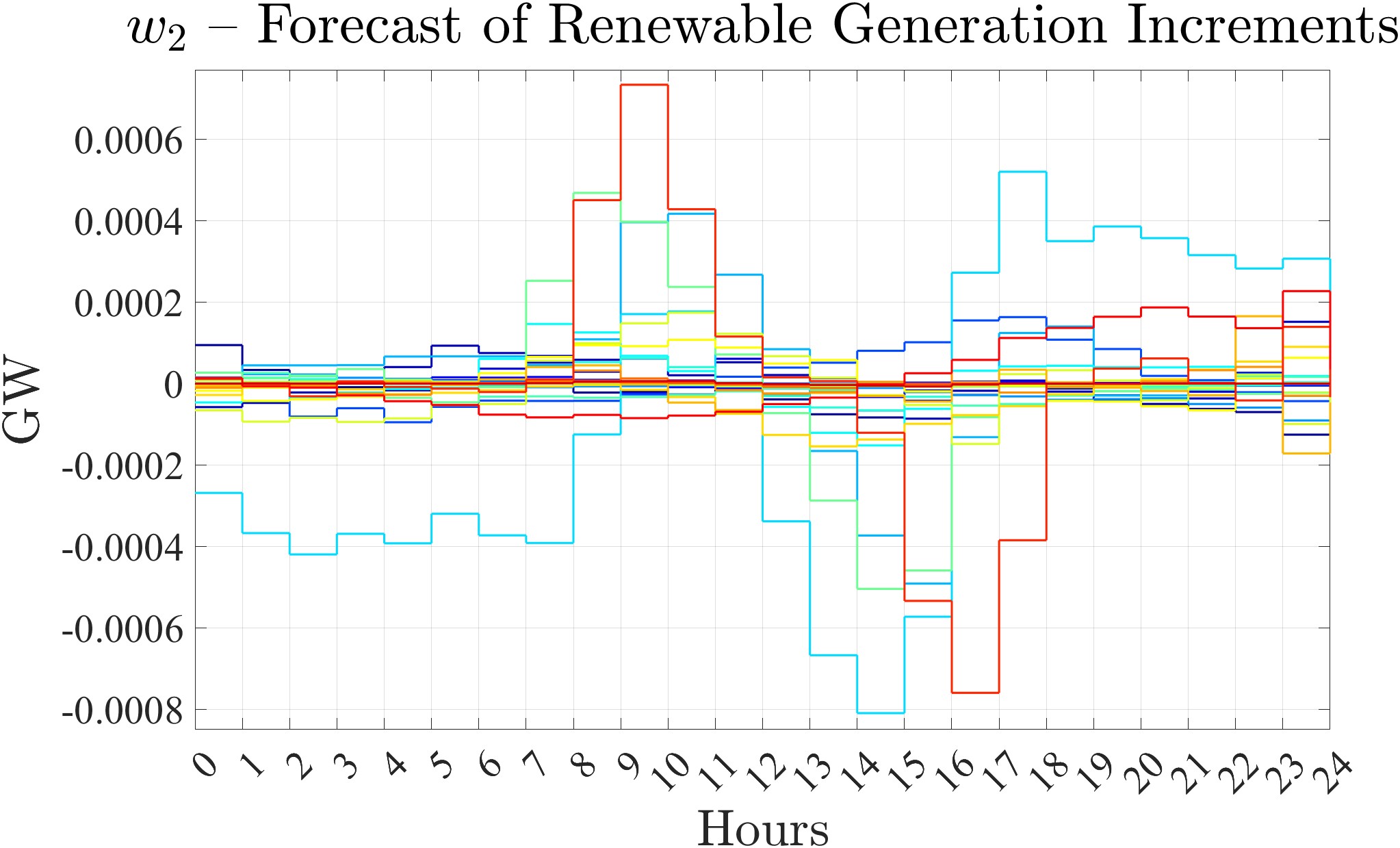}
		\end{subfigure} \hfill
	\end{minipage}%
	\hspace{0.03\linewidth}
	\begin{minipage}{0.11\textwidth}
		\includegraphics[width=\textwidth]{Legend}
	\end{minipage}
	\caption{Measurements and forecasts for the external signals.}
	\label{fig:delta}
\end{figure}

Table \ref{tab:param} reports the parameters used in the benchmark. Their selection is done according to the parameters used in similar simulation designs \cite{ bevrani_RobustPowerSystem_2014, ersdal_ModelPredictiveLoadFrequency_2016, kundur_PowerSystemStability_2022}. The sampling time of the systems is selected as $\tau = 2.5$ [s]. It follows that, for each hour, i.e.\ for each new data sample, 1440 steps of duration $\tau$ are considered in the control simulation.
\begin{table}[tbp!]
	\caption{Parameters in the EEA-ENB.}
	\centering
	\begin{tabular}{ccccccccc} 
		\hline
		$\tau$ & & $T_{\text{p},i}$ & & $K_{\text{p},i}$ && $\eta_{i}^\text{c}$ && $\eta_{i}^\text{d}$\\
		\hline
		2.5 [s] && 25 [s] && 0.05 $\left[\frac{\text{Hz}}{\text{GW}}\right]$ && 0.9 && 1.1 \\
		\hline
	\end{tabular}	
	\label{tab:param}
\end{table}
We use linear interpolation to compute the external inputs $\Delta P_i^{\text{load, meas}}(k)$ , for $k = 0, \ldots, 34559$, from the data of $P_i^{\text{load, meas}}(h)$ available every hour, for $h = 1,\ldots,24$.
The same approach is used for renewable generation measurements $P_i^{\text{ren, meas}}$, and for the forecasts $P_i^{\text{ren, for}}$, $P_i^{\text{ren, for}}$. The resulting signal variations are in Fig.\ \ref{fig:delta}.

\subsection{Comparison with other Benchmarks in the Field}
Several benchmarks for the simulation of power networks are present in the literature. Among the most popular ones, we can report the various implementations of IEEE buses \cite{_PowerEnergy}. Those benchmarks are oriented towards the simulation of power networks for electrical engineering applications. The benchmark we propose is instead oriented to the implementation of distributed control techniques, with predictive control as primary objective. To the best of our knowledge, an LFC-oriented benchmark modeled using data from the EEA is not present in the literature. A similar study was performed in \cite{ersdal_ModelPredictiveLoadFrequency_2016} for the simulation of the Northern European network, but without considering renewable generation and ESSs.

\subsection{Performance Metrics}
The performance of the control system are measured using the value of the quadratic cost function that will be formally introduced in the control problem \eqref{eq:MPC} of Section \ref{sec:control_MPC}. Since the EEA-ENB is structured for the implementation of distributed control techniques, the overall cost over one day of simulation for the network controlled with the centralized MPC architecture \eqref{eq:MPC} represents the optimality target for every alternative control formulation. 

Another performance indicator of the control strategy is the computation time required to obtain the control action. Specifically, centralized MPC might not be suited for real-time control of the EEA-ENB due to the excessive computation time required to obtain the optimal control action. Distributed architectures are usually faster in obtaining the control law since they distribute the computational burden among the control agents, but they are more complex to implement. 

{ As a part of the benchmark we provide the data for a control simulation of the network using the centralized MPC scheme \eqref{eq:MPC} for one day of operation of the network. Both the stage cost at each step, the overall cost for one day, and the total computation time are provided as a reference for alternative control strategies. Further details are in Section \ref{sec:discussion}. The end-user can consider these indexes as provided for a direct comparison, or perform their centralized MPC simulation with a different cost function to use personalized metrics.}

\subsection{Alternative Test Cases}
We propose the simulation for a single day of operation of the network. However, data is available at \cite{_ENTSOE} for every day of the year. Seasonality plays an important role in power generation from renewable sources. For example, solar production can increase or decrease depending on the presence of clouds, the temperature, and the length of the day. Load data is also affected by seasons. Evaluating the network with the average data about load and renewable generation of the four different seasons will give a clear view of the effectiveness of the control strategy considered over an entire year, with a mitigated variability introduced by a single day selection.

\subsection{Output Data}
Executing the benchmark will provide data about the electrical machine angles deviations, their frequency deviations, the energy stored in each area, the total power production and exchange with the ESSs, and the power transmitted over the tie lines. Those quantities are used to compute the performance metrics, and to evaluate the control strategy. Thus, both the evolution of the states of the system and the control actions can be collected and stored.

\subsection{Essential Properties}
The constraints provided for the frequency operation are essential for the stability of the network. Any prolonged deviation from the intervals provided will lead to emergency operation modes or failure of components, which in turn may generate cascading blackouts in the network.
The implementation of soft constraints on the state can allow for this deviation outside safety margins, but always considering the stability of the operation of the network and the economic cost of such deviations. For more information, we refer the reader to  \cite{ersdal_ModelPredictiveLoadFrequency_2016, haesalhelou_SurveyPowerSystem_2019}.

Regarding the MPC implementation, both the feasibility and stability properties should be met \cite{bemporad_RobustModelPredictive_1999, mayne_ConstrainedModelPredictive_2000}. Moreover, for the robustness of the system to the disturbances, which are the variations of the load and the renewable generation, an in-depth analysis of their evolution over an extended time window should be performed to characterize them correctly. Then, robust MPC synthesis methods could be used to guarantee this property \cite{bemporad_RobustModelPredictive_1999}. 
	
\section{Accessing the Benchmark} 

\subsection{Links to Sources, Limitations, Costs, and Licensing}
The benchmark is implemented in Matlab (r2023b), and the necessary files to execute it are available at \cite{riccardi_CodeUnderlyingPublication_2024b}. Gurobi Optimizer\footnote{\url{https://www.gurobi.com}} is required for the computation of the centralized MPC strategy. Alternatively, the Matlab Optimization Toolbox\footnote{\url{https://www.mathworks.com/products/optimization}} can be used with minimal modifications.

The benchmark is provided for free as it is under MIT the license. Data from \cite{_ENTSOE} used for the construction of the benchmark is publicly available {\color{black} under the Creative Commons Attribution 4.0 International License (CC-BY 4.0)}.
\subsection{Documentation}
Documentation for the benchmark is available at \cite{riccardi_CodeUnderlyingPublication_2024b}. In the following, we provide the user with information about the functions used. The data from \cite{_ENTSOE} has been preliminarily checked for integrity, replacing missing entries with linear interpolations, and reported on a consistent scale. This process was performed with specialized scripts reported in the online documentation. The resulting preprocessed data set is also part of the benchmark and provided as .csv files. The benchmark is constituted by the following files:
\begin{itemize}
	\item[--] {\tt main.m}: this is the principal script to run the control simulation.
	\item[--] {\tt data\_import.m}: {\color{black}this script reads the preprocessed data about load demands and renewable generation measurements and forecasts stored in .csv files, and returns the parameters and signals required for the simulation. }
	\item[--] {\tt state\_update\_network.m} and {\tt state\_update\_model.m}, which are identical files in this first formulation, but might be distinguished later to implement model mismatches or parameters inaccuracies. The former is used to simulate the system dynamics, and the latter as a prediction model for the centralized MPC strategy. 
	\item[--] {\tt objective\_function.m}: this function takes as inputs the parameters, the current value of the state, and the inputs and external inputs over the prediction window to return the total cost over that window.
	\item[--] {\tt plot\_results.m}: used to produce plots of the simulation results and input data.
\end{itemize}

\section{Discussion for Future Comparison}  \label{sec:discussion}
\subsection{Reference Approach: Centralized Predictive Control}
\label{sec:control_MPC}
The LFC problem has been extensively studied in the literature. As a source of references to existing approaches for its solution, we refer to the survey \cite{ranjan_LiteratureSurveyLoad_2022}. To provide a comparison case for the implementation of distributed control techniques, we have implemented a centralized MPC scheme \cite{ersdal_ModelPredictiveLoadFrequency_2016, parisio_ModelPredictiveControl_2014b}. Considering the system described in \eqref{eq:overall_dynamics}, we define at each time step $k$ the following centralized control problem:
\begin{equation}
	\begin{matrix}
		\displaystyle \min_{\bar{\mathbf{u}}} & \displaystyle \sum_{j=1}^{N} \mathbf{x}^\intercal(j|k) \mathbf{R} \mathbf{x}(j|k) + \mathbf{u}^\intercal(j-1|k) \mathbf{Q} \mathbf{u}(j-1|k) \hfill \\
		\text{s.t.} & \text{dynamics \eqref{eq:overall_dynamics}} \hfill \forall v_i \in \mathcal{V}  \\
		&  \mathbf{x}(0|k) = \mathbf{x}(k) \hfill\\
		& 			\text{constraints \eqref{eq:states_limits}} \hfill \forall v_i \in \mathcal{V}  \\
		& 			\text{constraints \eqref{eq:input_limits}} \hfill \forall v_i \in \mathcal{V}\\
	\end{matrix}
	\label{eq:MPC}
\end{equation}
were $\bar{\mathbf{u}} = \left[\mathbf{u}^\intercal(0|k) \cdots \mathbf{u}^\intercal(N-1|k)\right]^\intercal$, $\mathbf{R}$ and $\mathbf{Q}$ are diagonal cost matrices of appropriate dimensions such that, for each area $i$, $\mathbf{R}_i = \text{diag}[100;\,10;\,1]$ and $\mathbf{Q}_i = \text{diag}[1;\,1;\,1]$, and $N = 30$ is the prediction horizon. { For the external signals $\mathbf{w}$ we use the current measurement for $j = 1$, and their forecast for the remaining time steps.} According to the receding horizon logic of MPC, we apply $\mathbf{u}(0|k)$ to the system, discard the remaining control actions, and iterate. Problem \eqref{eq:MPC} is a quadratic optimization problem, hence efficient algorithms for solving it exist in the literature. For example, the problem can be solved using an active-set or an interior point algorithm. 
	
\begin{figure}[tbp!] 
	\begin{minipage}{.85\textwidth}
		\begin{subfigure}[t]{0.48\textwidth}
			\includegraphics[width=\textwidth]{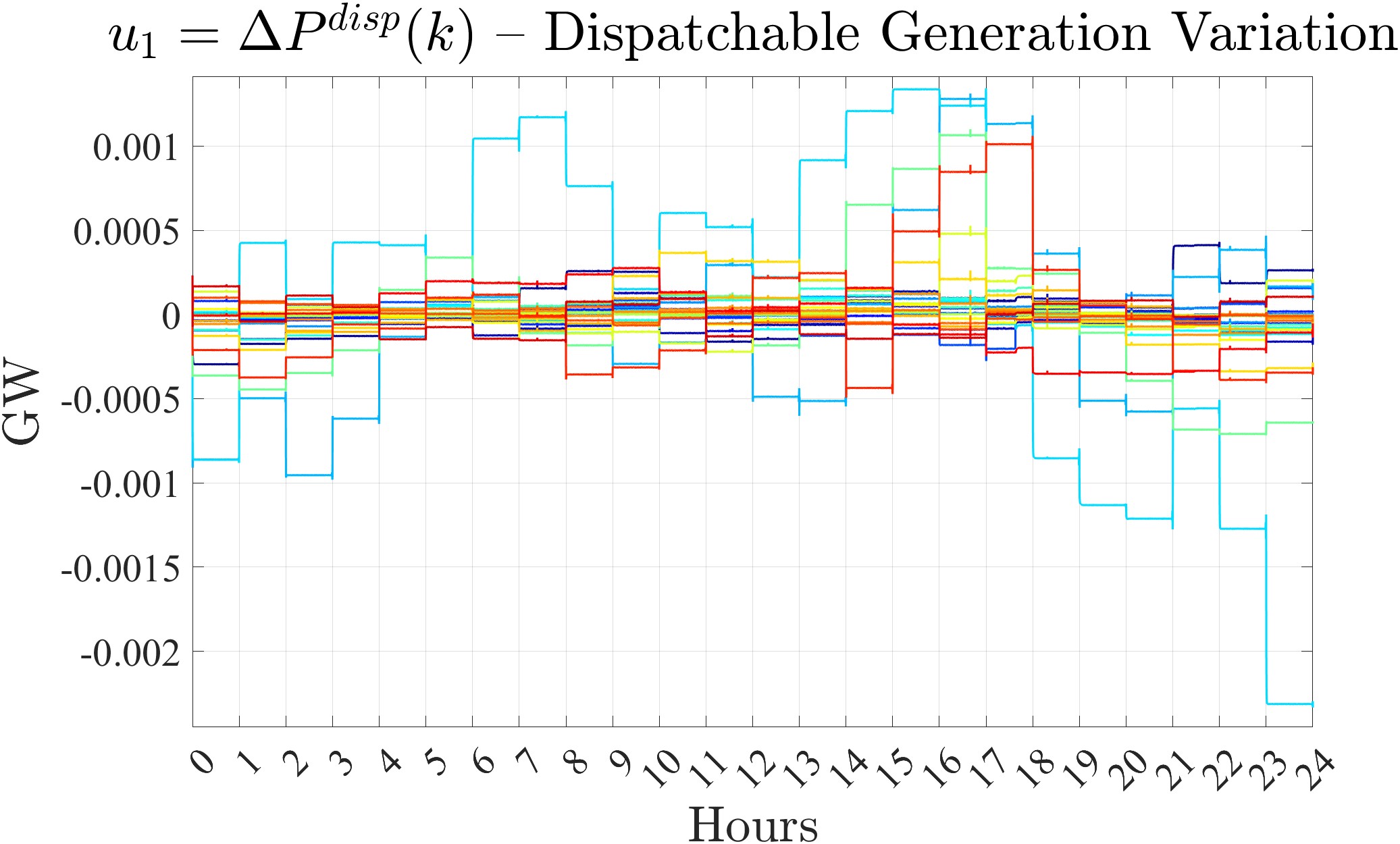}
			\caption{Input $u_1$.}
		\end{subfigure} \hfill
		\begin{subfigure}[t]{0.48\textwidth}
			\includegraphics[width=\textwidth]{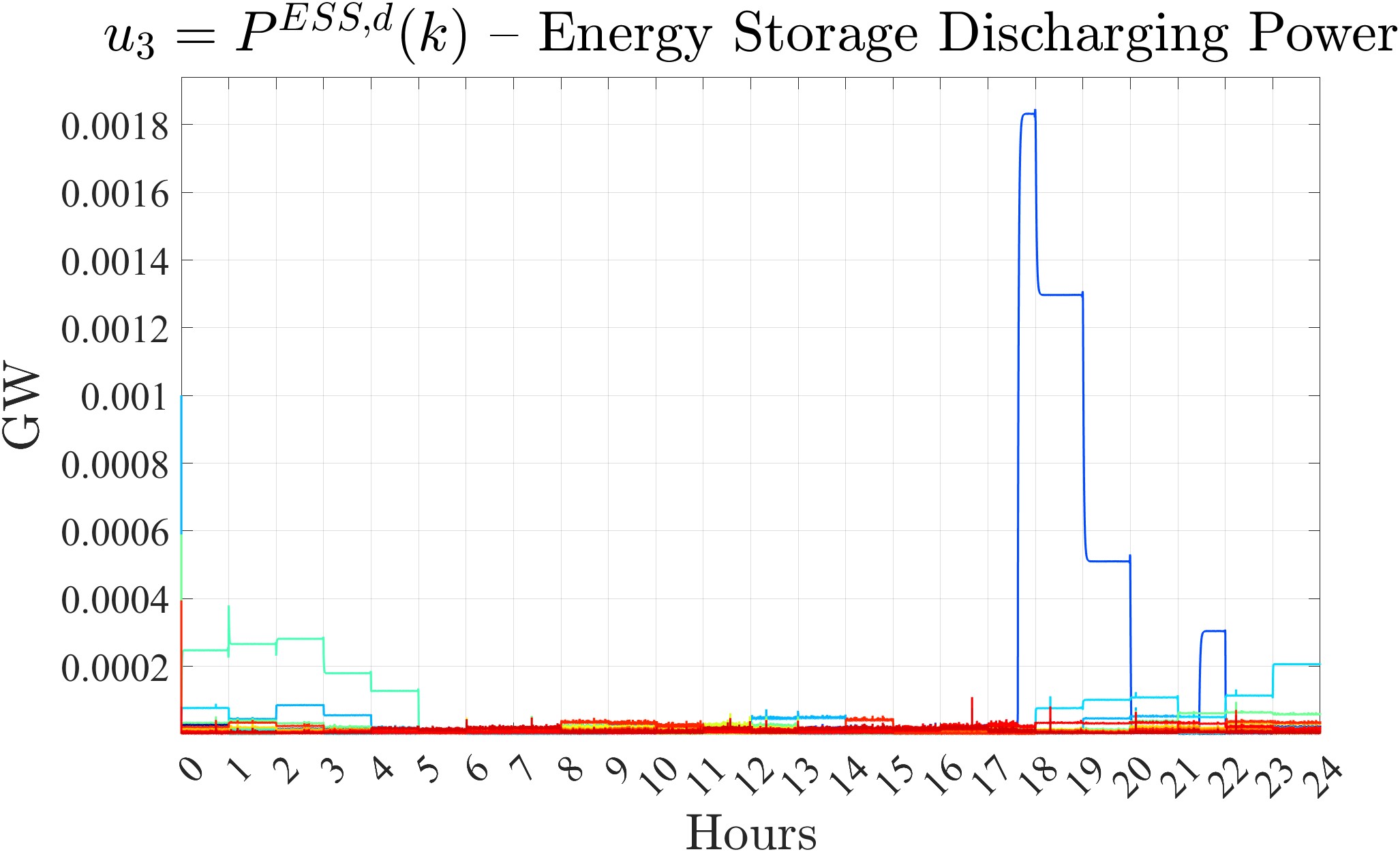}
			\caption{Input $u_3$.}
		\end{subfigure} \hfill
		\begin{subfigure}[t]{0.48\textwidth}
			\includegraphics[width=\textwidth]{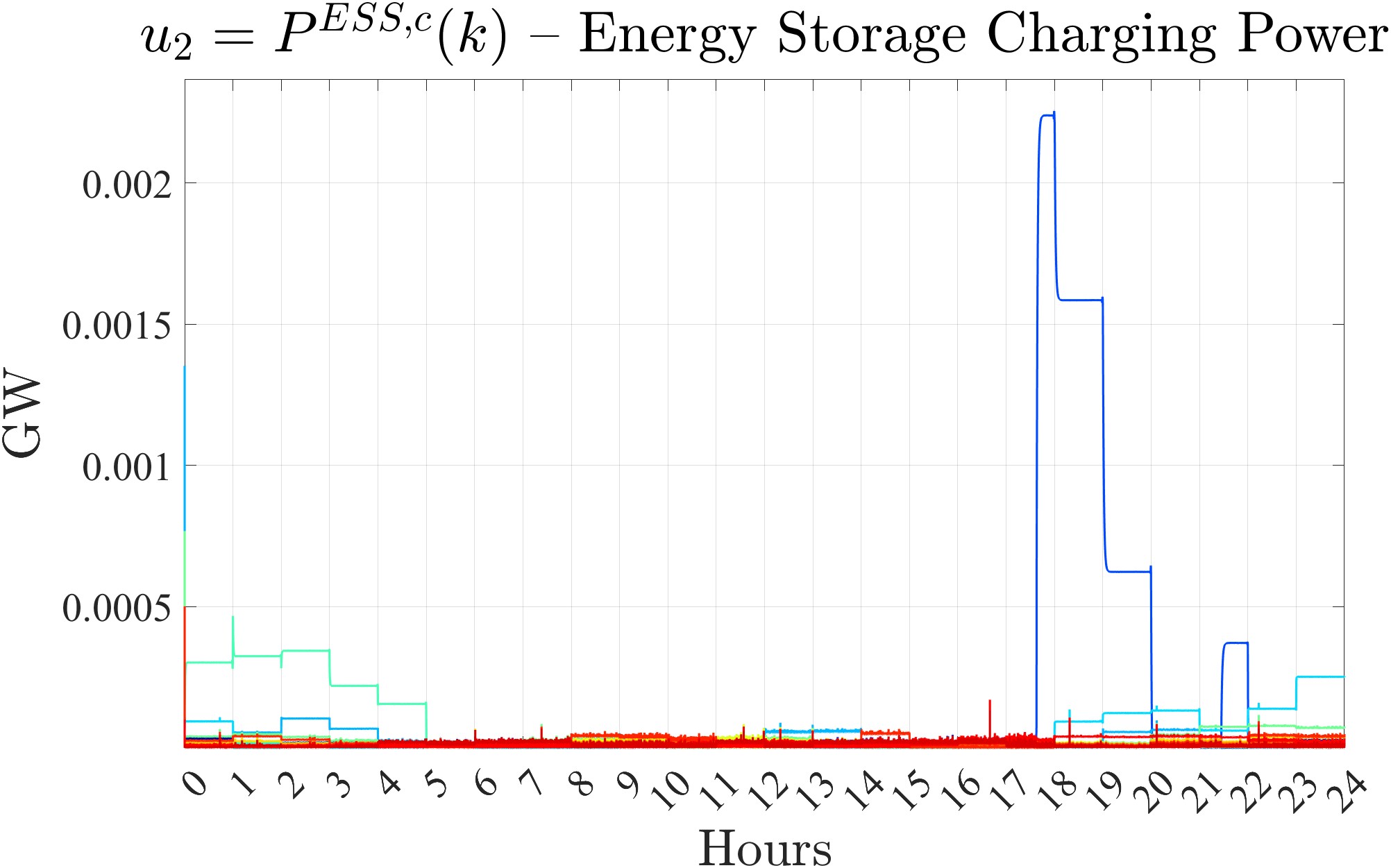}
			\caption{Input $u_2$.}
		\end{subfigure} \hfill
		\begin{subfigure}[t]{0.48\textwidth}
			\includegraphics[width=\textwidth]{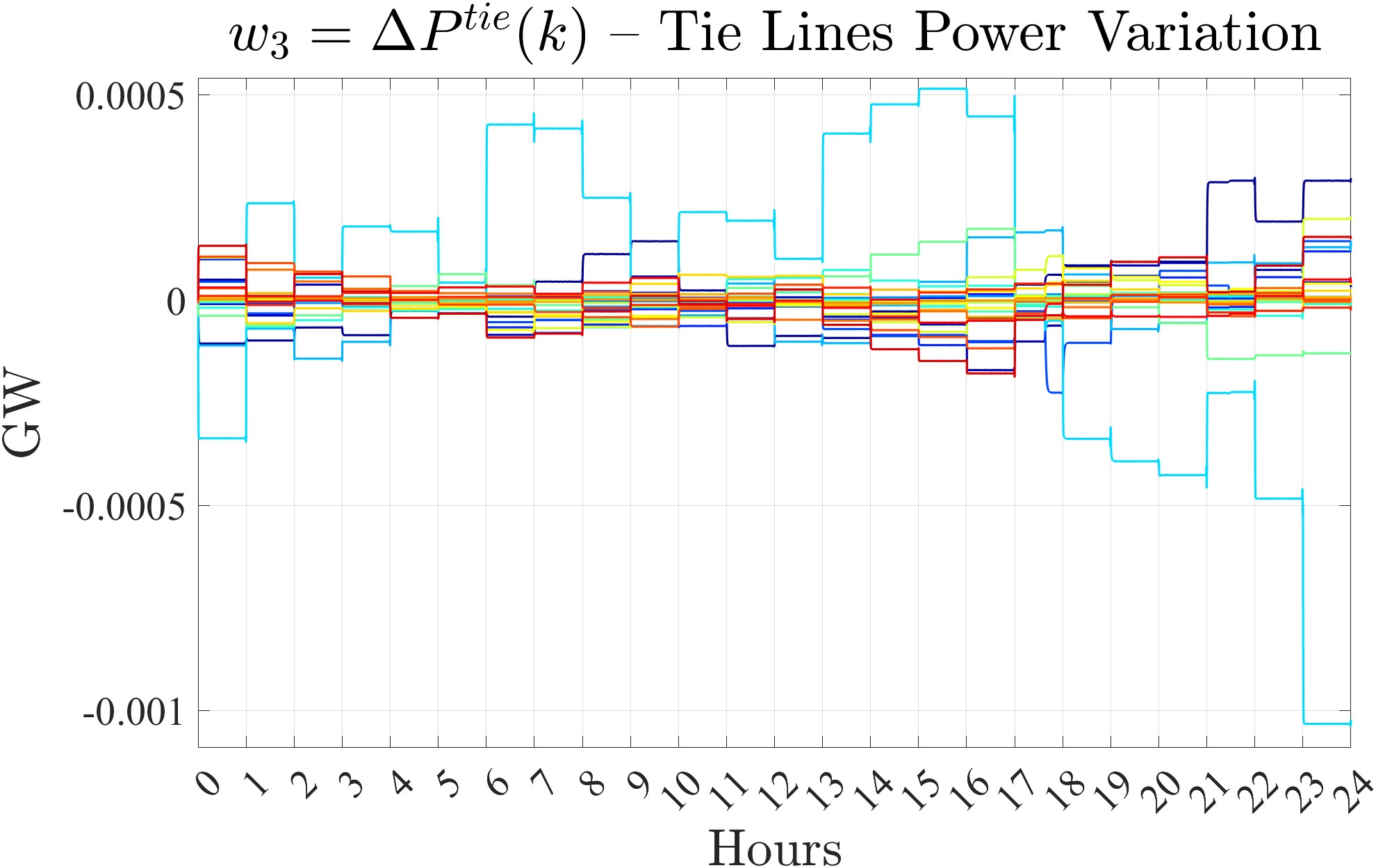}
			\caption{Power transmitted over the tie lines.}
		\end{subfigure} 
	\end{minipage}%
	\hspace{0.03\linewidth}
	\begin{minipage}{0.11\textwidth}
		\includegraphics[width=\textwidth]{Legend}
	\end{minipage}
	\caption{Evolution of the inputs and of the power transmitted over the tie lines.}
	\label{fig:inputs}
\end{figure}

For the benchmark, the optimization is performed with Gurobi Optimizer using the barrier algorithm. The simulation required 206 [h], 15[m], and 24[s]\footnote{ The amount of time required to run the simulation is only indicative, and serves to understand the complexity of the system. The end-user should re-run the benchmark on their own system to obtain results that are comparable with the architectures at study.} to be completed using a processor Intel Xeon E5-2637v3, with a base clock of 3.5 GHz, and coupled with 128 GB of RAM. The solution of the optimal control problem is the vector of inputs $\mathbf{u}(k)$, for $k = 0, ... , 34559$, (corresponding to 24 hours of simulation) reported in Fig.\ \ref{fig:inputs}.
In the same figure, the power transmitted over the tie lines connecting electrical areas is reported too, which allows quantifying the power trade necessary across electrical areas for the optimal operation of the network from a centralized and a cooperative perspective. Interactions among electrical areas, or agents in a generalized setting, is indeed one of the critical aspects of a distributed control strategy \cite{feleCoalitionalControlCooperative2017a, maestre_DistributedModelPredictive_2014}, and often one of the aspects characterizing the control strategy itself or the definition of the sub-networks in cooperation/competition. 

\begin{figure}[tbp!] 
	\begin{minipage}{.85\textwidth}
		\hfill
		\begin{subfigure}[t]{0.48\textwidth}
			\includegraphics[width=\textwidth]{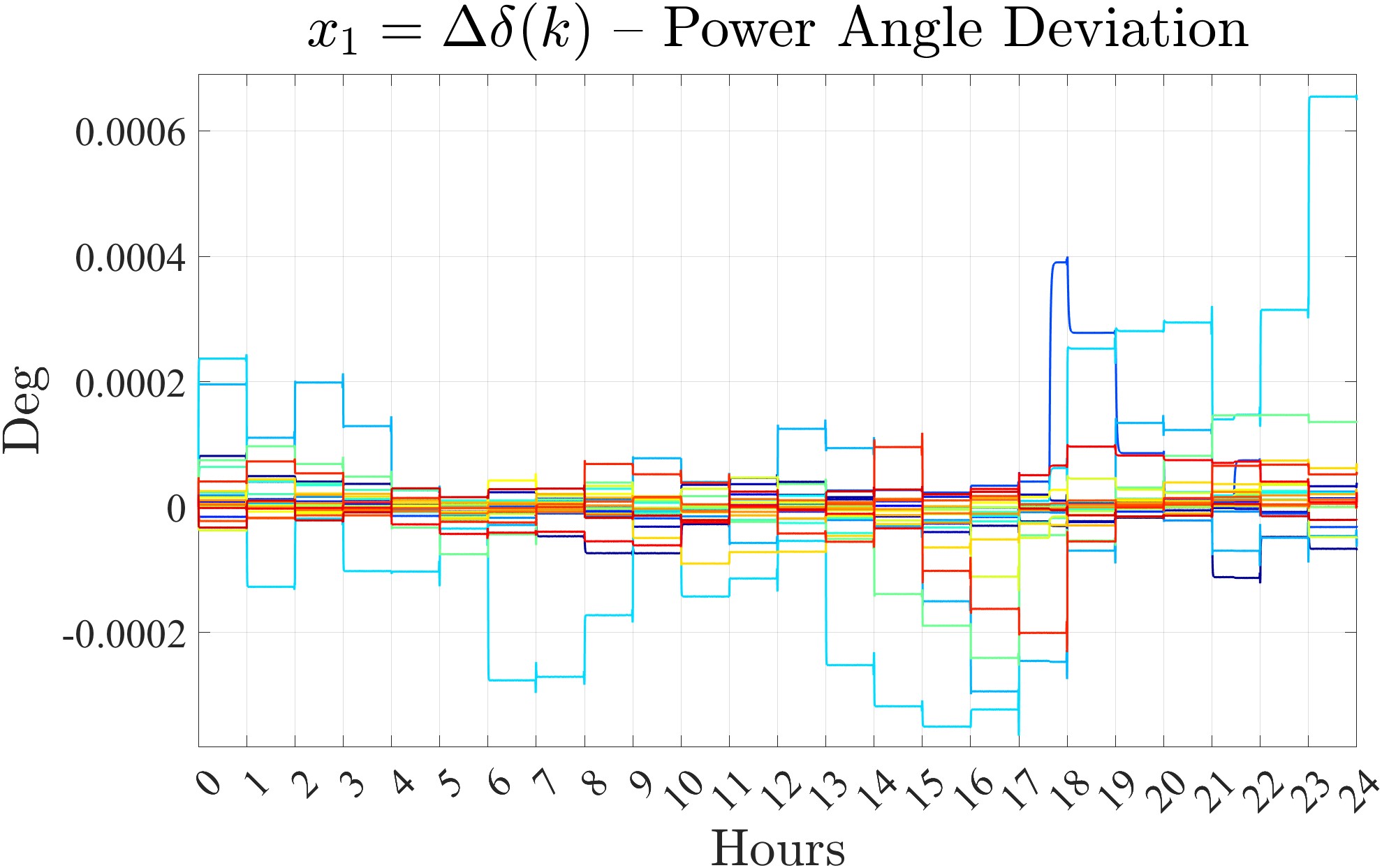}
			\caption{State $x_1$.}
		\end{subfigure} \hfill
		\begin{subfigure}[t]{0.48\textwidth}
			\includegraphics[width=\textwidth]{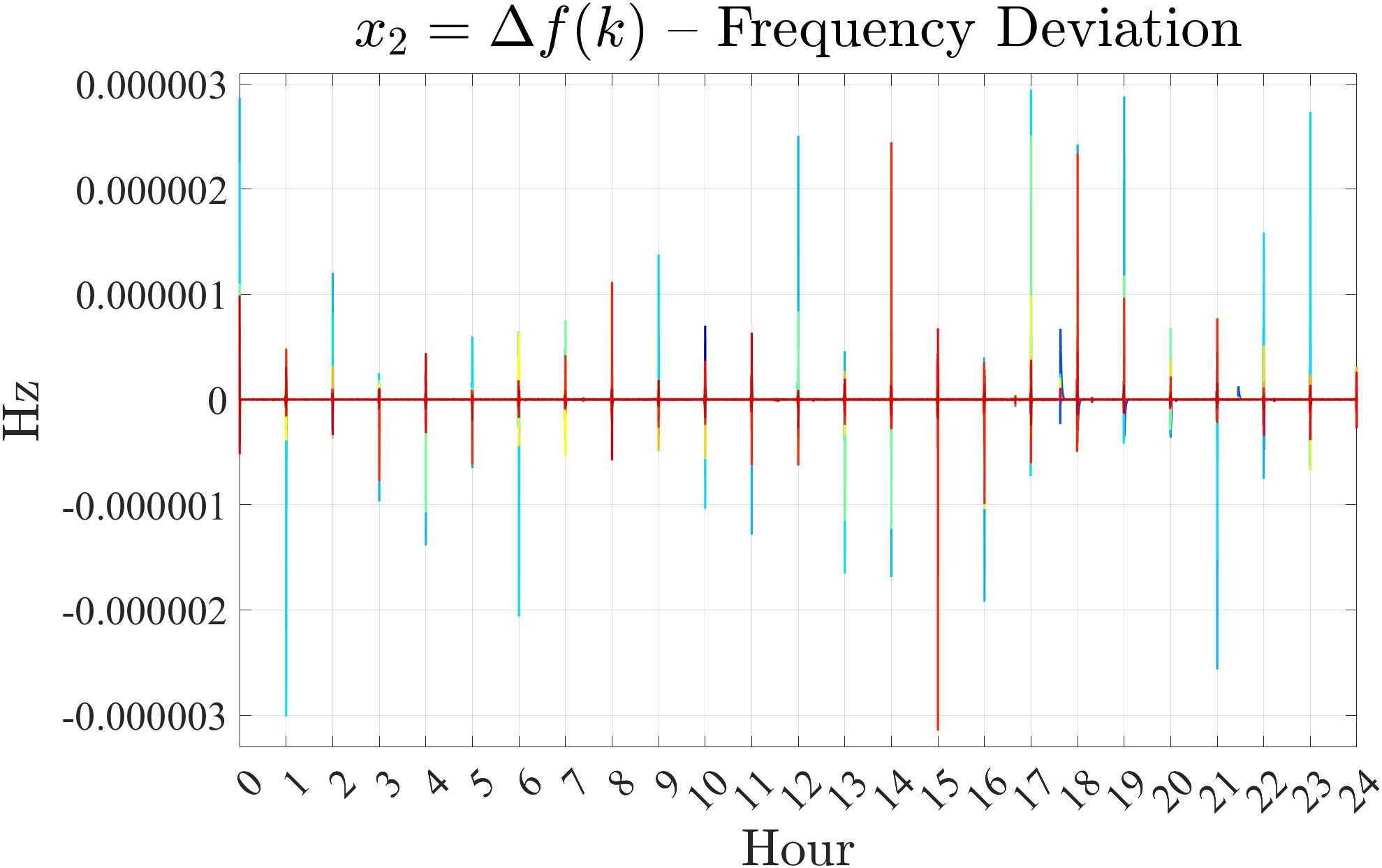}
			\caption{State $x_2$.}
		\end{subfigure} \hfill \\
		\hfill
		\begin{subfigure}[t]{0.48\textwidth}
			\includegraphics[width=\textwidth]{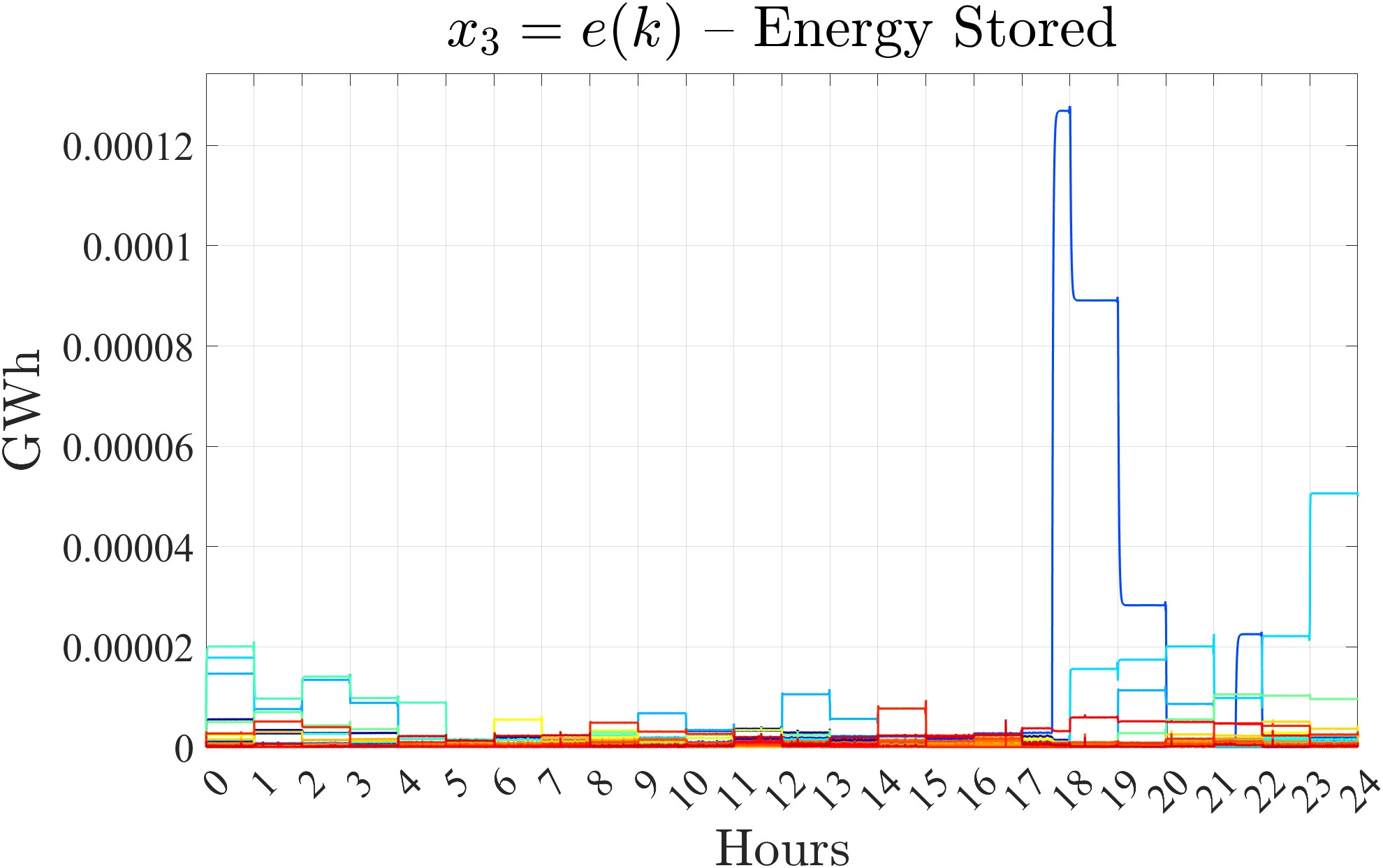}
			\caption{State $x_3$.}
		\end{subfigure} \hfill
		\begin{subfigure}[t]{0.48\textwidth}
			\includegraphics[width=\textwidth]{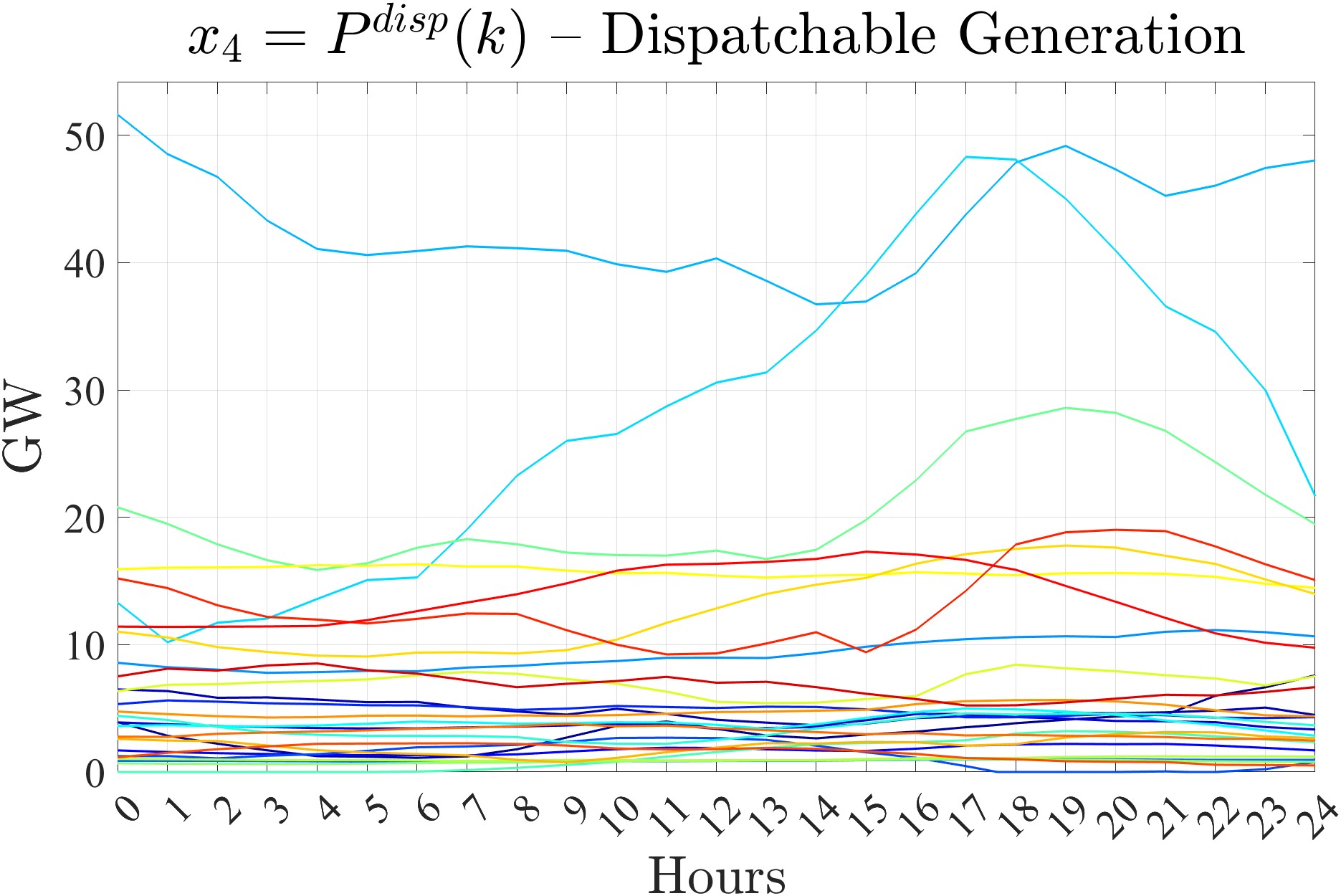}
			\caption{State $x_4$.}
		\end{subfigure} \hfill
	\end{minipage}%
	\hspace{0.03\linewidth}
	\begin{minipage}{0.11\textwidth}
		\includegraphics[width=\textwidth]{Legend}
	\end{minipage}
	\caption{Evolution of the states.}
	\label{fig:states}
\end{figure}

The evolution of the states resulting from the sequence of inputs obtained through MPC is reported in Fig.\ \ref{fig:states}. For the overall power balance of each country, please consult the online repository \cite{riccardi_CodeUnderlyingPublication_2024b}.

\begin{figure}[tbp!] 
	\centering
	\includegraphics[width=.45\linewidth]{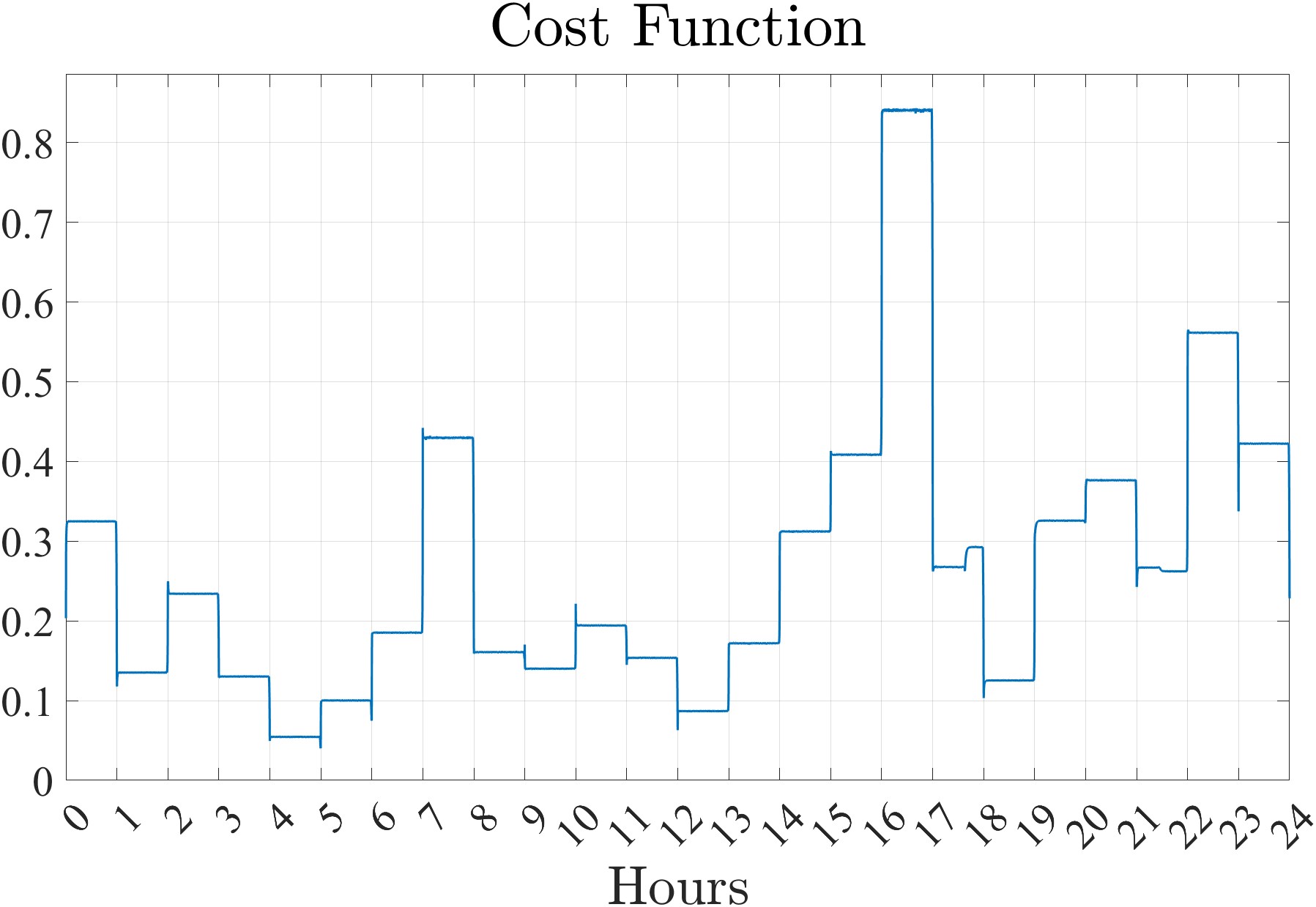}
	\captionof{figure}{\color{black}Evolution of the cost function.}
	\label{fig:cost}
\end{figure}

The evolution of the cost function is presented in Fig.\ \ref{fig:cost}. Its value is the metric to evaluate distributed control techniques w.r.t.\ the proposed centralized approach. In the application of distributed control, the performance of the system usually decreases to achieve auxiliary objectives such as improvement in the computation time required to obtain the control action, reduction in the volume of information exchanged across the network (with advantages in the sector of security too), or reliability and redundancy of control in the presence of faults or unforeseen events. 

\subsection{Other Possible Control Approaches}
To develop a deeper insight into the topics of LFC, MPC, and distributed control some references are reported in the following. 
In the LFC literature, see also \cite{ranjan_LiteratureSurveyLoad_2022}, many articles focus on PI control strategies where tuning of the parameters is performed through various optimization techniques \cite{gupta_LoadFrequencyControl_2021, kumar_WhaleOptimizationController_2020, raj_LoadFrequencyControl_2023}. Despite their promising validations and a general increase in performance w.r.t.\ conventional PI-based LFC, all these strategies still lack the fundamental advantages of model-based control: optimization of performance indices, incorporation of constraints into the control problem, ability to compensate for known external signals, and multi-objective optimization. On the other hand, PI control is easier to implement and has a faster computation speed.

A similar optimization-based tuning approach is used in \cite{elsisi_BatInspiredAlgorithm_2016}, this time to tune the parameters of an MPC controller. 

In \cite{ersdal_ModelPredictiveLoadFrequency_2016} a centralized MPC approach similar to the one proposed in \eqref{eq:MPC} is presented, but in \cite{ersdal_ModelPredictiveLoadFrequency_2016}, data of the Nordic transmission system is used to build an electricity network model. Power plant models are used to characterize the generation dynamics of each electrical area, but ESSs and renewable generation are not considered. Economic MPC is addressed too in \cite{ersdal_ModelPredictiveLoadFrequency_2016} and the results of centralized MPC are compared with a PD controller.

The use of hydro-pumped storage for LFC of microgrids including renewable generation is explored in  \cite{coban_LoadFrequencyControl_2022}, where a control architecture based on a decentralized PD controller tuned using a Quasi-Netwon optimization method is implemented. 

A characterization of an electricity network using hybrid dynamics is provided in \cite{parisio_ModelPredictiveControl_2014b}. There, PWA dynamics and binary decision variables are used, leading to a hybrid MPC problem formulated as a Mized Integer Linear Program (MILP), and solved using a branch-and-bound optimization algorithm. This approach is used to model three different aspects: the ESS, the operational mode of the microgrid, and the operational mode of the generation plants. 

In \cite{vlahakis_DistributedLQRDesign_2019} a distributed Linear Quadratic Regulator (LQR) is implemented to tackle the LFC problem. Methods for the distributed computation of the LQR control action are used in \cite{vlahakis_DistributedLQRDesign_2019} to increase the modularity and the scalability of the control architecture. The advantages of this approach rely on the ease of implementation, and on the low computation time needed to compute the control action. The disadvantages are that the applicability of the approach is limited to linear unconstrained control problems.

A way to address Economic MPC can be found in \cite{jia_CooperationBasedDistributedEconomic_2019}. There, both the LFC and economic load dispatch problems for power networks are considered. Those problems are usually approached using hierarchical control structures, where the economic load dispatch is at the upper level, and LFC at a lower level. Instead, in \cite{jia_CooperationBasedDistributedEconomic_2019} the two problems are considered simultaneously, to improve the economic performance of the systems. 

Another Economic MPC strategy is reported in \cite{golpira_FrameworkEconomicLoad_2014}. There, a multi-objective genetic algorithm predictive control technique is used to simultaneously optimize the conflicting objectives of LFC and security-constrained economic dispatch.

\subsection{Summary}
The LFC problem has been widely investigated in the literature, and we have proposed a benchmark to evaluate distributed control strategies for solving it. The challenges arising in recent years are often related to the ever-growing use of distributed energy sources which are inertia-less and can affect the frequency of the network. The use of ESSs can mitigate this effect, and we have proposed strategies to model their dynamics, extensions, and future directions for their exploration. {\color{black}Accordingly, future control strategies of electricity networks should account for the presence of distributed energy sources, and the implementation of distributed control strategies 
for efficient and resilient operation of the network.} Those aspects are indeed part of the centralized MPC formulation that we have proposed as a benchmark scheme. An efficient distributed control strategy is expected to perform worse than the centralized one proposed here if only the value of the cost function is considered, but has additional properties such as a reduced computation complexity and computation time, a lower shared volume of information, or enhanced privacy, security and resilience properties. 

\subsubsection*{Acknowledgments}
\noindent{This project has received funding from the European Research Council (ERC) under the European Union’s Horizon 2020 research and innovation program (Grant agreement No. 101018826) -- Project CLariNet}

\input{references}
\end{document}

%% file: references.tex
%
%
 \bibliographystyle{plain}
 \bibliography{references}